\documentclass[iop,apj,tighten]{emulateapj}

\usepackage[T1]{fontenc}
\usepackage{amssymb}
\usepackage{amsmath}
\usepackage{graphicx}
\usepackage{xspace}
\usepackage{natbib}
\usepackage{txfonts}
\usepackage{microtype}

\submitted{Accepted for Publication in The Astrophysical Journal}

\begin{document}

\title{The transient accreting X-ray pulsar XTE\,J1946$+$274:\\ 
Stability of the X-ray properties at low flux and updated orbital solution}

\author{
Diana M. Marcu-Cheatham\altaffilmark{1,2},
Katja Pottschmidt\altaffilmark{1,2},
Matthias K\"uhnel\altaffilmark{3},
Sebastian M\"uller\altaffilmark{3},
Sebastian Falkner\altaffilmark{3},
Isabel Caballero\altaffilmark{4},
Mark H. Finger\altaffilmark{5},
Peter J. Jenke\altaffilmark{6},
Colleen A. Wilson-Hodge\altaffilmark{7},
Felix F\"urst\altaffilmark{8},
Victoria Grinberg\altaffilmark{9},
Paul B. Hemphill\altaffilmark{10},
Ingo Kreykenbohm\altaffilmark{3},
Dmitry Klochkov\altaffilmark{11},
Richard E. Rothschild\altaffilmark{10},
Yukikatsu Terada\altaffilmark{12},
Teruaki Enoto\altaffilmark{13},
Wataru Iwakiri\altaffilmark{14},
Michael T. Wolff\altaffilmark{15},
Peter A. Becker\altaffilmark{16},
Kent S. Wood\altaffilmark{15},
J\"orn Wilms\altaffilmark{3}
}

\altaffiltext{1}{CRESST \& Department of Physics, University of Maryland Baltimore County,
  1000 Hilltop Circle, Baltimore, MD 21250, USA}

\altaffiltext{2}{NASA Goddard Space Flight Center, Astrophysics
  Science Division, Greenbelt, MD 20771, USA}

\altaffiltext{3}{Dr.\ Karl Remeis-Observatory \& ECAP, University
  Erlangen-Nuremberg, Sternwartstr.\ 7, 96049 Bamberg, Germany}

\altaffiltext{4}{Laboratoire AIM, CEA/IRFU, CNRS/INSU, Universit\'e
  Paris Diderot, CEA DSM/IRFU/SAp, F-91191 Gif-sur-Yvette}

\altaffiltext{5}{Universities Space Research Association, National
  Space Science and Technology Center, 320 Sparkman Drive, Huntsville,
  AL 35805, USA}

\altaffiltext{6}{University of Alabama in Huntsville, 301 Sparkman
  Drive, Huntsville, AL 35899, USA}

\altaffiltext{7}{Astrophysics Office, ZP 12, NASA Marshall Space
  Flight Center, Huntsville, AL 35812, USA}

\altaffiltext{8}{Cahill Center for Astronomy and Astrophysics,
  California Institute of Technology, Pasadena, CA 91125, USA}

\altaffiltext{9}{Massachusetts Institute of Technology, Kavli
  Institute for Astrophysics, Cambridge, MA 02139, USA}

\altaffiltext{10}{University of California, San Diego, Center for
  Astrophysics and Space Sciences, 9500 Gilman Dr., La Jolla, CA
  92093-0424, USA}

\altaffiltext{11}{Institut f\"ur Astronomie und Astrophysik,
  Universit\"at T\"ubingen (IAAT), Sand 1, 72076 T\"ubingen, Germany}
 
\altaffiltext{12}{Graduate School of Science and Engineering, Saitama
  University, 255 Simo-Ohkubo, Sakura-ku, Saitama City, Saitama
  338-8570, Japan}

\altaffiltext{13}{Department of Astronomy and The Hakubi Center for
  Advanced Research, Kyoto University, Kitashirakawa-Oiwake-cho,
  Sakyo-ku, Kyoto 606-8502, Japan}
 
\altaffiltext{14}{RIKEN Nishina Center, 2-1 Hirosawa, Wako, Saitama
  351-0198, Japan }

\altaffiltext{15}{Space Science Division, Naval Research Laboratory,
  Washington, DC, USA}

\altaffiltext{16}{School of Physics, Astronomy, and Computational
  Sciences, MS 5C3, George Mason University, 4400 University Drive,
  Fairfax, VA, USA}
                                   
\begin{abstract}
  We present a timing and spectral analysis of the X-ray pulsar
  XTE\,J1946$+$274 observed with \textsl{Suzaku} during an outburst
  decline in 2010 October and compare with previous
  results. XTE\,J1946$+$274 is a transient X-ray binary consisting of
  a Be-type star and a neutron star with a 15.75\,s pulse period in a
  172\,d orbit with 2--3 outbursts per orbit during phases of
  activity. We improve the orbital solution using data from multiple
  instruments. The X-ray spectrum can be described by an absorbed
  Fermi-Dirac cutoff power law model along with a narrow Fe K$\alpha$
  line at 6.4\,keV and a weak Cyclotron Resonance Scattering Feature
  (CRSF) at $\sim35$\,keV. The \textsl{Suzaku} data are consistent
  with the previously observed continuum flux versus iron line flux
  correlation expected from fluorescence emission along the line of
  sight. However, the observed iron line flux is slightly higher,
  indicating the possibility of a higher iron abundance or the
  presence of non-uniform material. We argue that the source most
  likely has only been observed in the subcritical (non-radiation
  dominated) state since its pulse profile is stable over all observed
  luminosities and the energy of the CRSF is approximately the same at
  the highest ($\sim5\times10^{37}$\,erg\,s$^{-1}$) and lowest
  ($\sim5\times10^{36}$\,erg\,s$^{-1}$) observed 3--60\,keV
  luminosities.
\end{abstract}


\section{INTRODUCTION}\label{sec:intro}

The X-ray pulsar \objectname{XTE\,J1946$+$274} was discovered during a
three-month long outburst in 1998 September by the All-Sky Monitor
(ASM) on the \textsl{Rossi X-Ray Timing Explorer} (\textsl{RXTE})
\citep{smith:98}. Pulsations with a period of 15.83\,s were first
detected by \citet{ wilson:98} using data from the Burst And Transient
Source Experiment (BATSE) on board the \textsl{Compton Gamma-Ray
  Observatory} (\textsl{CGRO}). XTE\,J1946$+$274 was found to be a
High Mass X-ray Binary (HMXB) with a Be IV/IVe stellar companion
\citep{verrecchia:02}. \citet{wilson:03} determined an orbital period
of 169.2\,days, an orbital inclination of $\sim46^{\circ}$, and a
distance of $9.5\pm2.9$\,kpc using \textsl{RXTE} and BATSE data.
Between 1998 and 2001, XTE\,J1946$+$274 experienced an outburst
approximately every half-orbit: \citet{campana:99} observed periodic
flaring of the X-ray source repeating every $\sim$80\,days. Between
1999 September and 2000 July, the outbursts were monitored with the
Indian X-ray Astronomy Experiment (IXAE) and the data were analyzed by
\citet{paul:01}. \citet{paul:01} and \citet{wilson:03} presented pulse
profiles with double-peaked structures.

The strong magnetic field ($\sim10^{12}$\,G) of the neutron star
enforces collimated accretion along the field lines and quantizes the
electron energy states perpendicular to those field lines. When X-ray
photons in the column interact through resonant scattering with these
quantized electrons they produce an absorption-line-like feature
observed in the spectrum at the energy
\begin{equation}
  E \approx \frac{11.56\,\mathrm{keV}}{1+z}
  \left(\frac{B_\mathrm{NS}}{10^{12}\mathrm{G}}\right)
\label{ecycl}
\end{equation}
where $B_\mathrm{NS}$ is the surface magnetic field, and $z$ is the
gravitational redshift, which is $\sim0.3$ for typical neutron star
parameters, and a line-forming region close to the surface. This is
known as a Cyclotron Resonance Scattering Feature (CRSF), which, as
can be seen in equation~(\ref{ecycl}), can be used to determine the
magnetic field strength of highly magnetized pulsars. The first
spectral analysis of XTE\,J1946$+$274 was performed by
\citet{heindl:01} using pointed \textsl{RXTE} data from the first
observed outburst in 1998. They found evidence for a CRSF with a
centroid energy of $\sim$36\,keV corresponding to a $B$-field of
$3.1(1+z)\times 10^{12}$\,G.

\begin{figure}
  \includegraphics[width=\columnwidth]{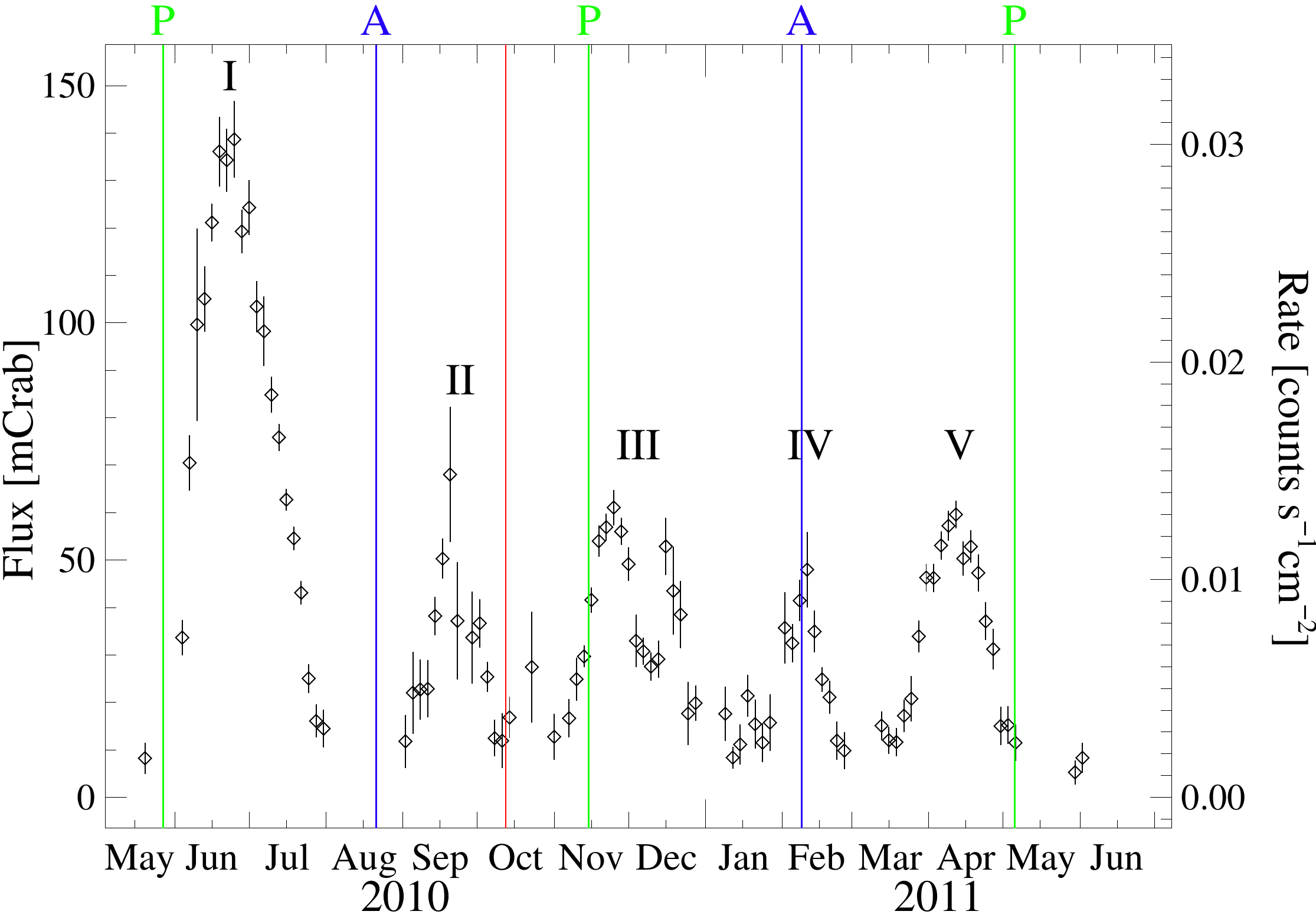}
  \caption{\textsl{Swift}-BAT 15--50\,keV XTE\,J1946$+$274 light curve
    of the series of outbursts in 2010--2011 with a binning of 3\,d,
    showing all bins with S/N$\gtrsim$2; the vertical red line
    represents the time of the \textsl{Suzaku} observation. The
    apastron (blue lines marked with ``A'') and periastron (green
    lines marked with ``P'') times were determined with the new
    orbital solution (see \S\ref{subsec:orbit}). The outbursts are
    marked I--V. The data were obtained from
    http://swift.gsfc.nasa.gov/results/transients/.}\label{fig:batlc}
\end{figure}

After 2001 October the source was quiescent until 2010 June. Starting
2010 June 4 the Burst Alert Telescope (BAT) on board of \textsl{Swift}
and the Gamma-ray Burst Monitor (GBM) on board of \textsl{Fermi}
observed a new strong outburst \citep{finger:10,krimm:10}. The BAT
light curve (Figure~\ref{fig:batlc}) shows that this $\sim$140\,mCrab
outburst was followed by four outbursts at about half the flux at
intervals of approximately 82, 75, 73, and 57 days. This behavior is
similar to that observed by \citet{campana:99} for the 1998--2001
outburst series.
 
\citet{caballero:10} found no sign of the CRSF at 35\,keV in a
preliminary analysis of \textsl{INTErnational Gamma-Ray Astrophysics
  Laboratory (INTEGRAL)} data of the first 2010 outburst. Using
\textsl{RXTE} and \textsl{INTEGRAL} data from the first outburst in
2010 June--July and \textsl{Swift}, \textsl{RXTE}, and
\textsl{INTEGRAL} data from the third outburst in 2010
November--December, \citet{muller:12} reported the possible presence
of a CRSF at 25\,keV (1.81$\sigma$ significance).

An iron (Fe) K$\alpha$ fluorescent line at 6.4\,keV is present in the
spectra. \citet{muller:12} reported a correlation between the Fe
K$\alpha$ line flux and the 7--15\,keV continuum flux.

In this paper we present a temporal and spectral analysis of
\textsl{Suzaku} data taken during the end of the second 2010 outburst
(red line in Figure~\ref{fig:batlc}) that allows for a spectral
analysis at the lowest flux to date. Due to its high broad-band X-ray
sensitivity and its imaging capability, \textsl{Suzaku} is an ideal
instrument for analyzing broad-band spectra and spectral features
(iron lines and CRSFs) for sources at very low fluxes. A first
temporal and spectral analysis of the same 2010 \textsl{Suzaku} data
was conducted by \citet{maitra:13} who reported the presence of a
broad CRSF at $\sim$38\,keV\footnote{Note that \citet{maitra:13} quote
  the resonance energy of a pseudo-Lorentzian line shape, the energy
  of the minimum of the line shape that is comparable to the CRSF
  energy values quoted elsewhere in this paper is $\sim$40\,keV
  \citep[see page~94 of][and \citealt{enoto:08a}]{mihara:95}.}. The
analysis we present here differs significantly from theirs, regarding
the spectral analysis itself as well as the breadth of the discussion.
The differences between our modeling choices are further explained in
\S\ref{sec:spec}. The CRSF width of $\sim$9\,keV found by
\citet{maitra:13} is rather broad and could indicate a contribution to
modeling the continuum \citep[for a demonstration of this effect
  see][]{muller:13}. In addition the source is not consistently
detected above 38\,keV in all spectral bins, even if broadly rebinned
\citep[large uncertainties have also been noted by][]{maitra:13}. This
is also the reason why we, contrary to \citet{maitra:13}, do not
conduct a pulse phase resolved analysis of the CRSF parameters. Though
not excluded, the 38\,keV line is thus an unlikely CRSF candidate. As
we show in \S\ref{sec:spec} there is a possibility that a less broad
line is present at $\sim$35\,keV instead.

The 2010--2011 outburst series was also monitored by
\textsl{Fermi}-GBM. Together with the available \textsl{RXTE},
\textsl{Swift}, and \textsl{Suzaku} data, these observations allow us
to refine the orbit parameters.

In \S\ref{sec:obs} we describe the \textsl{Suzaku} data and the data
reduction procedure, and provide an overview of the additional
multi-instrument data used in our analysis. In \S\ref{sec:timing} we
first examine the \textsl{Suzaku} light curves and hardness ratios. We
then determine the local pulse period and the energy resolved pulse
profiles which we compare with those observed with \textsl{RXTE}-PCA
during the bright first outburst of 2010. Last, but not least, we
present the improved orbital solution. In \S\ref{sec:spec} we present
the broad-band \textsl{Suzaku} spectral analysis. In
\S\ref{sec:discussion} and \S\ref{sec:summary} the results are
discussed and summarized, respectively.

\section{OBSERVATIONS AND DATA REDUCTION}\label{sec:obs}

\begin{figure*}
  \includegraphics[width=\columnwidth]{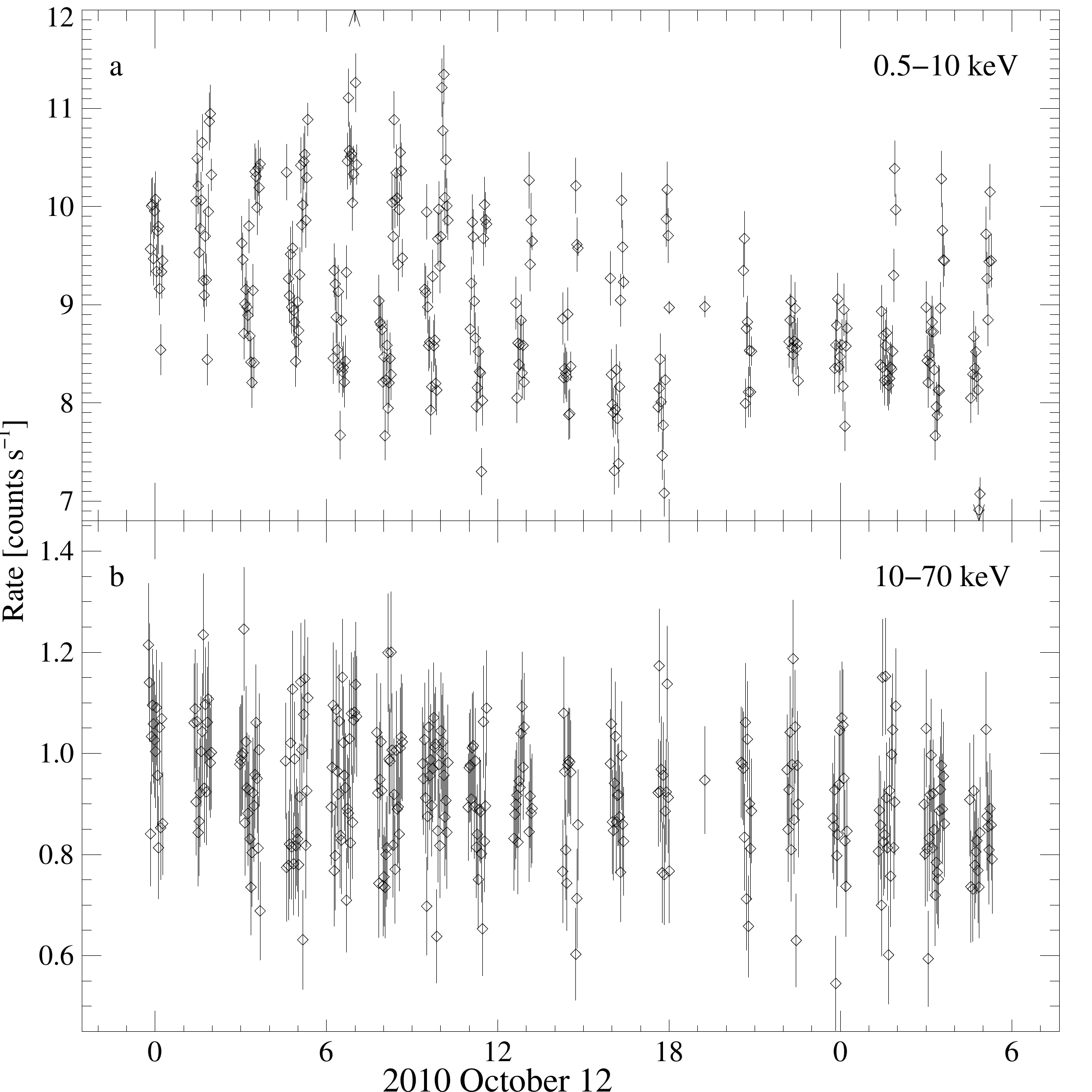}
\hfill
  \includegraphics[width=\columnwidth]{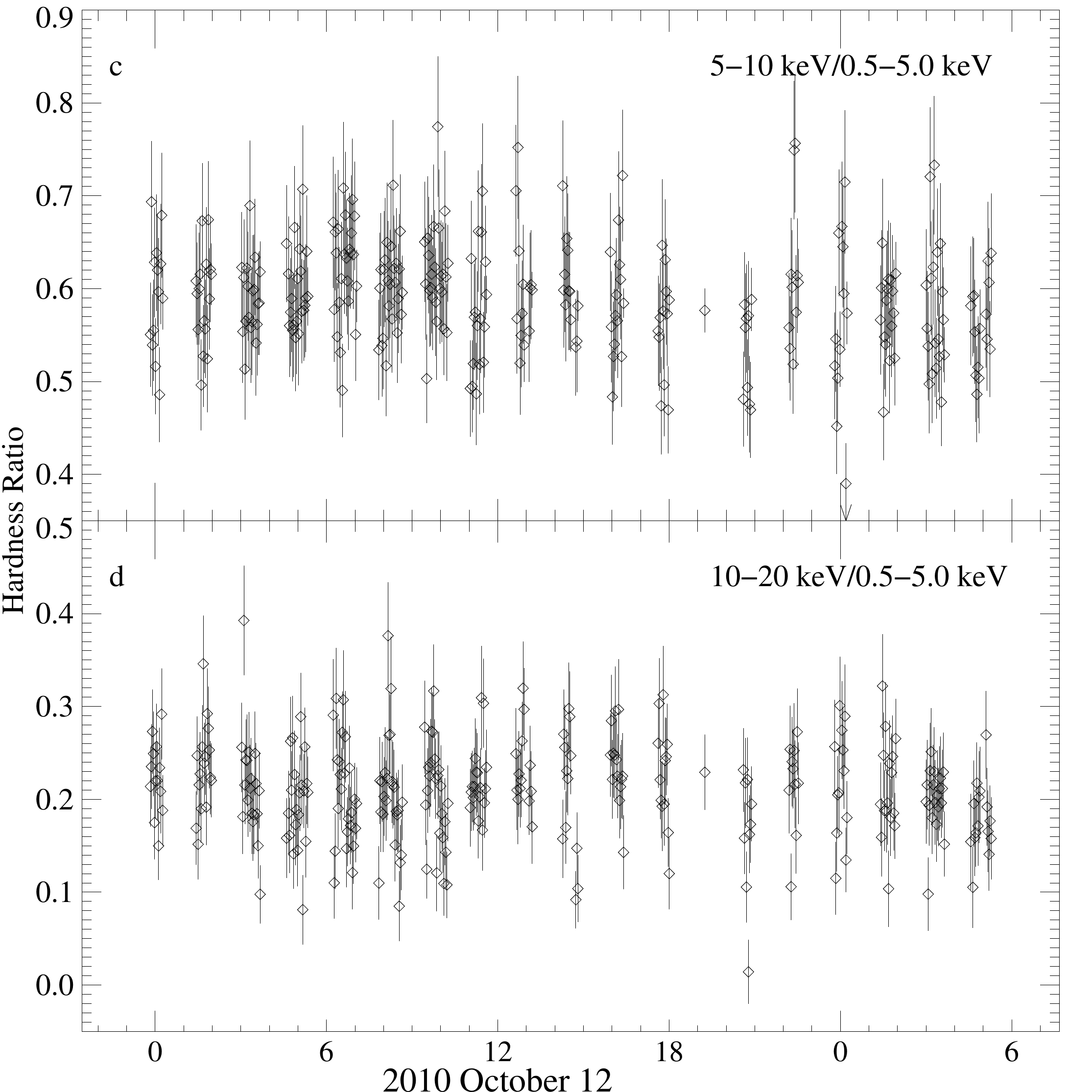}
  \caption{(a) Light curve of summed XIS 0, 1, and 3 count rates
    (0.5--10\,keV). (b) Dead-time corrected PIN light curve
    (10--70\,keV). Both light curves are background subtracted and
    binned to 128\,s. (c) Hardness ratio evolution for count rates in
    the energy bands 5--10\,keV and 0.5--5\,keV using XIS\,3. (d)
    Hardness ratio evolution for count rates in the energy bands
    10--20\,keV and 0.5--5\,keV using PIN and
    XIS\,3.}\label{fig:lc_hardness}
\end{figure*}

We study a $\sim$50\,ks \textsl{Suzaku} observation that occurred on
2010 October 11--13 (ObsID 405041010), during a minimum
between the second and third outburst of the 2010 outburst series,
when the 15--50\,keV flux was $\sim$10\,mCrab. We extracted data
obtained with the X-ray Imaging Spectrometer
\citep[XIS,][]{koyama:07}, and the PIN instrument from the High X-ray
Detector \citep[HXD,\,][]{takahashi:07}. The three functional units of
the XIS (CCD cameras 0, 1, and 3) were operated in the 1/4 window mode
during the observation in order to reduce pile up.  Data from the
Gadolinium Silicate Crystals (GSO, also part of HXD) were excluded due
to the weakness of the source above 40\,keV.

We reprocessed the XIS and PIN data and extracted data products
following the \textsl{Suzaku} Data Reduction (or ABC) Guide
\citep{abc_guide}. The reprocessing was performed using
\texttt{aepipeline}, applying the newest calibration as well as
standard data screening (with the default screening criteria). This
was done based on the \texttt{HEASOFT} v6.13 software package and the
calibration database (\texttt{CALDB}) releases HXD-20110913,
XIS-20130305 and XRT-20110630. We further filtered the screened XIS
events in order to exclude times of telemetry saturation. The events
for both XIS and PIN were transferred to the barycenter of the solar
system with \texttt{aebarycen}.

Using \texttt{xselect}, we first extracted XIS images, to which we
applied an attitude correction with \texttt{aeattcor2}, which further
corrects the attitude data for thermal wobbling using mean event
positions as a function of time. After comparing the images obtained
with and without applying \texttt{aeattcor2}, we concluded that the
additional attitude correction does not improve the moderate
systematic attitude instability which is visible in the images through
an elongated and double-peaked point spread function \citep[PSF,
  see][for further discussions of this
  effect]{suzaku_memo2010_06,suzaku_memo2010_05,suzaku_memo2010_04}.
As we will show in \S\ref{subsec:lc_hardness}, the systematic attitude
wobble has negligible effect on the spectral shape.

XIS source and background event files, light curves and spectra were
produced using \texttt{xselect} after selecting the extraction regions
in the XIS image. For bright sources, this step involves the
determination of possibly existing pile-up using \texttt{pileest}. For
XTE\,J1946$+$274 the pileup fraction was $<4\%$ in the center of the
PSF, thus the source was not bright enough to cause strong pile-up
during this observation. We used the same source extraction region for
the three XIS units and the two editing modes alternately used for
event storage (``$3\times3$'' and ``$5\times5$''): a circle with a
radius of 120\,pixels ($124\farcs8$) centered on the PSF. The circle
is large enough to contain most of the source events but not larger
than the window. The background regions were circles with radii of
95\,pixels ($98\farcs8$), located within the windows, but as far from
the PSFs as possible. XIS\,0 has a strip of unusable, masked pixels
near the edge of the detector and therefore our XIS\,0 background
region additionally avoided this zone \citep{suzaku_memo2010_10}.

The XIS source and background light curves were extracted with 128\,s
resolution in the energy bands 0.5--5\,keV, 5--10\,keV, and
0.5--10\,keV. Since the orbital period of the neutron star (172\,d)
is significantly larger than the duration of the observation (50\,ks),
we did not perform a binary star orbit correction. The XIS spectra
were binned to a resolution close to the half-width half-maximum of
the spectral resolution of the instrument \citep{nowak:12}. To
generate the energy and ancillary responses we used the
\texttt{xisrmfgen} and \texttt{xissimarfgen} tools, respectively. The
exposure time for each XIS CCD is $\sim$50\,ks, while the average
source count rates are $\sim$3.05\,counts\,s$^{-1}$ for XIS\,0,
$\sim$2.80\,counts\,s$^{-1}$ for XIS\,1, and
$\sim$3.48\,counts\,s$^{-1}$ for XIS\,3.

For PIN we applied energy filtering (10--20\,keV, 20--40\,keV,
40--70\,keV and 10--70\,keV) to the event files obtained after running
\texttt{aepipeline}, after which we extracted light curves with
\texttt{hxdpinxblc} with a time binning of 128\,s. This tool produces
the total dead-time corrected PIN light curve, the non X-ray
background light curve, and the background-subtracted source light
curve. We used \texttt{hxdpinxbpi} for the PIN spectral extraction
which provides the dead-time corrected PIN source spectrum and the Non
X-ray Background (NXB) and Cosmic X-ray Background (CXB)
spectra. Approximately 5\% of the PIN background are CXB and the
corresponding spectrum is simulated based on the description by
\citet{boldt:87}. The NXB light curve and spectrum produced by the
extraction tools are based on modeled events available for each
individual
observation\footnote{\url{ftp://legacy.gsfc.nasa.gov/suzaku/data/background/pinnxb\_ver2.0\_tuned/
 2010\_10/ae405041010\_hxd\_pinbgd.evt.gz}}. For the spectral modeling
we used the summed NXB and CXB background. The appropriate response
file for the specific calibration epoch was chosen
(\texttt{ae\_hxd\_pinhxnome9\_20100731.rsp}). For the PIN spectra we
applied a binning of a factor of 2 for the energy range
34--40\,keV. The exposure time for PIN is $\sim$43\,ks, while the
total average source count rate is
$\sim0.90\,\mathrm{counts}\,\mathrm{s}^{-1}$.

In addition to these \textsl{Suzaku} data we also used
XTE\,J1946$+$274 data from other instruments. The pulse profile
comparison in \S\ref{subsec:pulse} presents the \textsl{Suzaku}-XIS
and \textsl{Suzaku}-PIN data together with \textsl{RXTE}-PCA data from
the peak of the first outburst in 2010. The orbit determination in
\S\ref{subsec:orbit} is based on the complete 2010 outburst series.
The majority of pulse period measurements is provided by the
\textsl{Fermi}-GBM Pulsar
Project\footnote{http://gammaray.nsstc.nasa.gov/gbm/science/pulsars/}
while also including \textsl{Suzaku}-PIN, all available
\textsl{RXTE}-PCA, and \textsl{Swift}-XRT data. In
\S\ref{sec:discussion} we compare \textsl{Suzaku} results with results
from \citet{heindl:01} and \citet{muller:12} obtained with
\textsl{RXTE}, \textsl{Swift}, and \textsl{INTEGRAL}. For all
observations used in our analysis, the instruments that performed
them, their observation times, and their exposure times are listed in
Table~\ref{table:obs}.

\begin{deluxetable}{lcc} 
\tablecolumns{3} 
\tablewidth{\columnwidth} 
\tablecaption{Observations}
\tablehead{ 
\colhead{Satellite}  & \colhead{Observation} &
\colhead{Number of observations} \\
\colhead{Instrument}  & \colhead{time} &
\colhead{Total exposure time}}

\startdata
\textsl{Fermi} & 2010 Dec 16 -- 2011 May 1  & monitoring \\
GBM & $1^\mathrm{st}$ -- $5^\mathrm{th}$ outbursts in 2010  & \\
\hline\\
\textsl{RXTE} & 1998 Sept\,16 -- 1998 Oct\,14  & 12 observations \\
PCA, HEXTE$^{a}$ & $1^\mathrm{st}$ outburst in 1998 & $\sim$30\,ks \\
\hline\\
\textsl{RXTE} & 2010 Jun 20 -- 2010 Jul 16 & 17 observations \\
PCA$^{b}$ & $1^\mathrm{st}$ outburst in 2010 & $\sim$60\,ks \\
\hline\\
\textsl{RXTE} & 2010 Nov 23 -- 2010 Dec 07 & 9 observations \\
PCA$^{b}$ & $3^\mathrm{rd}$ outburst in 2010  & $\sim$23\,ks \\
\hline\\
\textsl{Swift} & 2010 Nov 26 -- 2010 Dec 28 & 8 observations \\
XRT$^{b}$ & $3^\mathrm{rd}$ outburst in 2010 & $\sim$16\,ks \\
\hline\\
\textsl{INTEGRAL} & 2010 Jun 20 -- 2010 Nov 30 & 5 observations \\
ISGRI$^{b}$ & $1^\mathrm{st}$ \& $4^\mathrm{th}$ outbursts in 2010 & $\sim$150\,ks \\
\hline\\
\textsl{Suzaku} & 2010 Oct 11--13 & 1 observation \\
XIS, PIN$^{c}$ & end of $2^\mathrm{nd}$ outburst in 2010 & $\sim$50\,ks
\enddata 
\tablecomments{$^a$ \citet[][their Table~1]{heindl:01}; $^b$ \citet[][first sentence of notes on their Table~1]{muller:12}; $^c$ this
  work (\S\ref{sec:obs}).}
\label{table:obs}
\end{deluxetable} 

\section{TEMPORAL ANALYSIS}\label{sec:timing}

\subsection{\textsl{Suzaku} Light Curves and Hardness Ratios}\label{subsec:lc_hardness}

Figures~\ref{fig:lc_hardness}a and \ref{fig:lc_hardness}b show the
background subtracted light curves for the summed count rates of XIS
0, 1 and 3 (0.5--10\,keV) and for the PIN count rate (10--70\,keV),
respectively. According to the \textsl{Swift}-BAT light curve in
Figure~\ref{fig:batlc}, the \textsl{Suzaku} observation was performed
at the end of a decreasing long-term flux trend. This appears to be
consistent with the PIN light curve, which might show a moderate
decline from $1.06\pm0.03\,\mathrm{counts}\,\mathrm{s}^{-1}$ in the
first satellite orbit of the observation to
$0.86\pm0.03\,\mathrm{counts}\,\mathrm{s}^{-1}$ in the last one. There
are no significant flares or dips observed. The XIS light curve
displays jumps between two count rate levels for most \textsl{Suzaku}
orbits. The effect can be observed in all three XIS units
individually, and it is consistent with the systematic attitude
instability mentioned in \S\ref{sec:obs}. This is aggravated by the
HXD aim-point used for this observation, since it is slightly
off-center on the XIS chips.

Figures~\ref{fig:lc_hardness}c and \ref{fig:lc_hardness}d show
hardness ratio evolutions for count rates in the energy bands
5--10\,keV and 0.5--5\,keV and for count rates in the energy bands
10--20\,keV and 0.5--5\,keV, respectively. We observe little structure
related to the systematic attitude instability in the XIS-PIN band
ratios, and no structure in the XIS-XIS band ratios. Since there are
no significant source related flux or hardness changes over the
observation, we do not perform a time resolved spectral analysis but
model the observation averaged spectra in \S\ref{sec:spec}.

\subsection{Pulse Period and Pulse Profiles}\label{subsec:pulse}

\begin{figure}
  \includegraphics[width=\columnwidth]{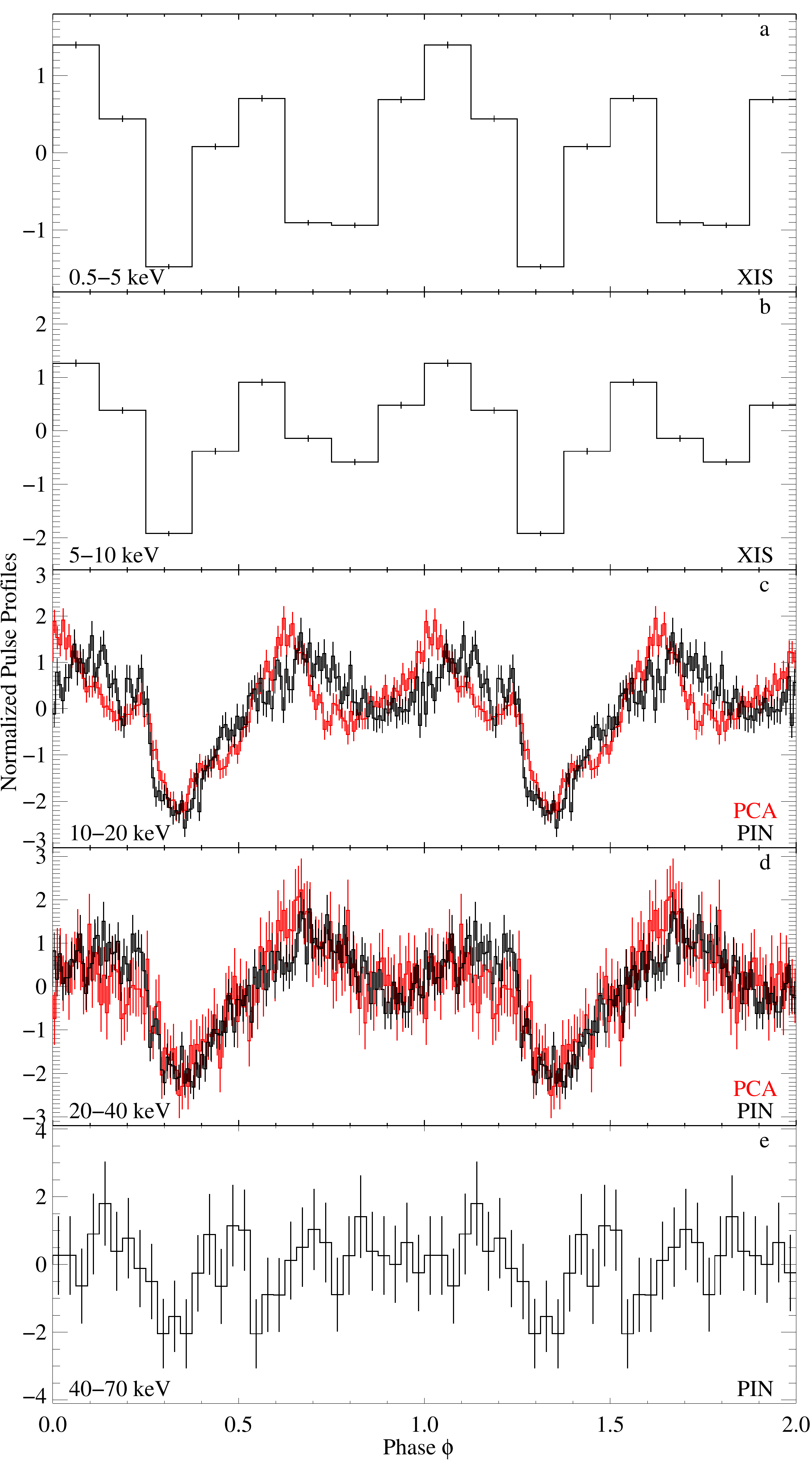}
  \caption{Energy resolved \textsl{Suzaku} and \textsl{RXTE} pulse
    profiles for the respective instruments and energy rages: (a)
    XIS\,3, 0.5--5\,keV, (b) XIS\,3, 5--10\,keV, (c) PIN (in black)
    and PCA (in red), 10--20\,keV, (d) PIN and PCA, 20--40\,keV, (e)
    PIN, 40--70\,keV.  The \textsl{RXTE}-PCA pulse profiles are from
    the peak of the bright first outburst in 2010 June (see
    \S\ref{subsec:pulse} for further discussion). The number of phase
    bins for XIS, PIN and PCA is 8, 128, and 128, respectively, with
    the exception of 32 for the 40--70\,keV PIN range. The period
    values the \textsl{Suzaku} and \textsl{RXTE} events were folded on
    are 15.750025\,s (this work) and 15.764\,s \citep[][]{muller:12},
    respectively. The profiles were normalized to show standard
    deviations above the mean.}\label{fig:pulse_profile}
\end{figure}

The XIS has a time resolution of 2\,s when in 1/4 window mode, while
the PIN has a resolution of 61\,$\mu$s \citep{tech_description}.
Therefore, only the PIN data were used for the pulse period
determination. Applying epoch folding
\citep{leahy:83,schwarzenberg-czerny:89a} to the screened,
barycenter-corrected, non-background subtracted PIN events, in the
10--40\,keV range, we determined a local pulse period of
15.750025(27)\,s. The uncertainty was estimated using Monte Carlo
light curve simulations as described in \S\ref{subsec:orbit}.

Based on this period and a reference time of MJD\,55481.714 for
phase\,0, we obtained pulse profiles in several energy bands by
folding the screened, barycenter-corrected events using 8 phase bins
for XIS (0.5--5\,keV and 5--10\,keV; note that the \textsl{Suzaku}-XIS
pulse profiles presented by \citet{maitra:13} are oversampled) and 128
phase bins for PIN (10--20\,keV, 20--40\,keV and
40--70\,keV). Figure~\ref{fig:pulse_profile} shows that up to 40\,keV
the pulse profiles are consistent in general structure: they are
double-peaked, with a deep ($\phi\sim0.35$) and a shallow minimum
($\phi\sim0.9$). In the 10--20\,keV range an additional narrow peak
feature is visible ($\phi\sim0.2$) before the deep minimum. The
shallow minimum is deeper at energies $\lesssim$5\,keV than at higher
energies. Similar behavior was found by \citet{wilson:03} during two
outbursts observed with \textsl{RXTE}-PCA in 1998 and 2001. We
determined the pulse fractions measured with PIN as the difference
between the maximum and minimum count rates of the profiles normalized
by mean count rate, and obtained values of $1.02\pm 0.09$ and $1.04\pm
0.12$ for the 10--20\,keV and 20--40\,keV energy ranges,
respectively. \citet{wilson:03} found pulsed fractions as high as 0.74
in the 2--30\,keV range during low-flux outbursts in 2001. No
pulsations are visible in the 40--70\,keV \textsl{Suzaku} profile.

Figures~\ref{fig:pulse_profile}c and \ref{fig:pulse_profile}d include
a comparison for the 10--20\,keV and 20--40\,keV energy bands between
the \textsl{Suzaku}-PIN pulse profiles from 2010 October 12 (end of
the second outburst) and the \textsl{RXTE}-PCA pulse profiles from
2010 June 26 (ObsID 95032-12-02-00, peak of the first outburst).  The
latter were obtained using the same light curve extraction criteria as
\citet{muller:12} used for the full PCA energy band and applying epoch
folding with the local period of 15.764\,s determined by their
analysis. This comparison emphasizes that the shapes of the profiles
obtained from the two instruments are very similar, especially at
higher energies, despite the large difference in flux:
\begin{align*}
 \mbox{10--20}\,\mathrm{keV \,flux}: &
 \begin{cases}
    1.57\,\times\,10^{-9}\,\mathrm{erg}\,\mathrm{s}^{-1}\,\mathrm{cm}^{-2} & \mathrm{PCA},\\
    2.10\,\times\,10^{-10}\,\mathrm{erg}\,\mathrm{s}^{-1}\,\mathrm{cm}^{-2} & \textsl{Suzaku},
 \end{cases}\\
 \mbox{20--40}\,\mathrm{keV \,flux}: & \begin{cases}
    1.12\,\times\,10^{-9}\,\mathrm{erg}\,\mathrm{s}^{-1}\,\mathrm{cm}^{-2} &  \mathrm{PCA},\\
    1.30\,\times\,10^{-10}\,\mathrm{erg}\,\mathrm{s}^{-1}\,\mathrm{cm}^{-2} & \textsl{Suzaku}.
  \end{cases}
\end{align*}
The \textsl{Suzaku} fluxes were derived from the spectral best fit
model presented in \S\ref{subsec:model} and the \textsl{RXTE}-PCA
fluxes from the spectral best fit of the averaged observations during
the peak of the first outburst \cite[epoch 1 fit of][]{muller:12}.

\subsection{Orbit Determination}\label{subsec:orbit}

\begin{figure}
  \includegraphics[width=\columnwidth]{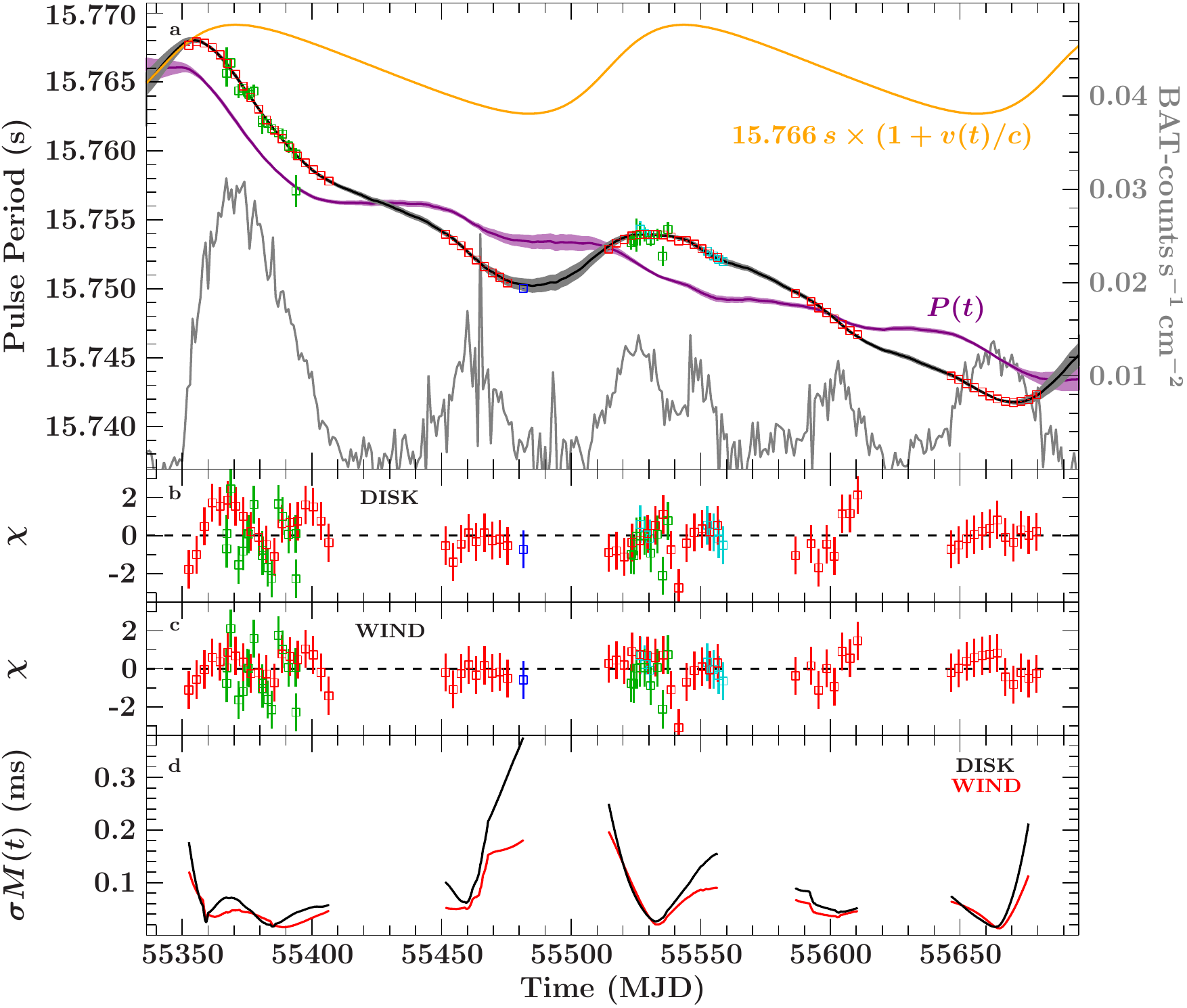}
  \caption{Orbit determination: The upper panel (a) shows the observed
    evolution of barycenter corrected pulse period values obtained
    with \textsl{Fermi}-GBM (red squares), \textsl{Suzaku}-PIN (dark
    blue square), \textsl{RXTE}-PCA (green squares), and
    \textsl{Swift}-XRT (light blue squares). It also shows the overall
    modeled pulse period evolution (black), the modeled intrinsic spin
    period evolution (purple), and the orbital motion effect (orange)
    for the DISK model. The BAT 15--50\,keV light curve is
    overplotted in gray. The lower panels show the residuals for
    fitting (b) the DISK model and (c) the WIND model to the observed
    evolution. Both models include intrinsic and orbital effects, but
    differ in the choice of the luminosity exponent $\alpha$. The
    model uncertainties are taken into account in the residuals and in
    the overall model and intrinsic spin period evolution drawn as a
    band in lighter colors. Panel (d) shows the Monte Carlo simulation
    of the model uncertainties: As described in the text the
    \textsl{Swift}-BAT light curve was randomized within its
    uncertainties, resulting in a different best fit of
    equation~(\protect\ref{eq:periodobs}) to the pulse periods during
    each run. The standard deviation of all calculated pulse period
    evolutions at the times where period measurements are available is
    shown for different assumptions of $\alpha$ in black (DISK) and
    red (WIND). These values are interpreted as model uncertainties
    for the final fits.}\label{fig:gbm_profile}
\end{figure}

The observed pulse period over time as measured by an observer is due
to the intrinsic spin-up or spin-down of the neutron star, caused,
e.g., by accretion torques, and on due to the Doppler shift by orbital
motion. Usually the Doppler shift dominates changes in the measured
pulse period. For XTE\,J1946+274, however, the neutron star undergoes
a strong spin-up during outbursts such that the orbital parameters of
the system could not be constrained well in the past.

\citet{wilson:03} were able to describe the pulse frequencies as
measured by \textsl{CGRO}-BATSE and \textsl{RXTE}-PCA during the
outburst series between 1998 and 2001 using a piece-wise linear
approximation of the intrinsic spin-up. Their best fit with a reduced
$\chi^2$ of $\chi^2_\mathrm{red}=5.94$ for 37 degrees of freedom
(d.o.f.) shows that this simplified approximation cannot give a good
description of the measured period evolution.

In the most simple model for the angular momentum transfer of the
infalling material onto the neutron star \citep{ghosh:79b}, the period
change of the neutron star is connected to the luminosity $L$ via
\begin{equation}\label{eq:pdotrelation}
 -\dot{P} \propto P^2 L^\alpha
\end{equation}
where $\alpha=1$ for wind and $\alpha=6/7$ for disk accretion.
Assuming that the luminosity of the source is proportional to the
measured flux $F$, the pulse period at the time $t$ is then given by
\begin{equation}\label{eq:periodtime}
 P(t) = P_0 + a (t - t_0) - b \int_{t_0}^{t}
 \left(\frac{P(t')}{P_0}\right)^2
 \left(\frac{F(t')}{F_\mathrm{ref}}\right)^\alpha d{t'}
\end{equation}
where $P_0$ is the pulse period at the reference time, $t_0$, $b$ is
the torque strength, and $F_\mathrm{ref}$ is a reference flux. The
model also takes a constant spin-change, $a$, into account, which
could be caused, e.g., the propeller effect \citep{illarionov:75}. We
obtain the observed pulse period $P_\mathrm{obs}(t)$ by applying the
Doppler shift caused by the orbital motion to $P(t)$ as defined in
equation~(\ref{eq:periodtime}):
\begin{equation}\label{eq:periodobs}
 P_\mathrm{obs} = P(t) (1 + v(t)/c)
\end{equation}
where $v(t)$ the orbital velocity of the neutron star projected on the
line of sight and where $c$ is the speed of light. The orbital
parameters needed to calculate $v(t)$ are the orbital period,
$P_\mathrm{orb}$, the time of periastron passage, $\tau$, the
projected semi-major axis, $a_\mathrm{sm} \sin i$, where $i$ is
the inclination, the eccentricity $e$, and the longitude of periastron
$\omega$, such that
\begin{equation}\label{eq:radvel}
 v(t) = \frac{2 \pi a_\mathrm{sm} \sin i}{P_\mathrm{orb} (1-e^2)^{1/2}} \left(\cos(\theta(t)+\omega) + e \cos \omega\right)
\end{equation}
where $\theta(t)$ is the true anomaly found by solving Kepler's
equation, which itself depends on the orbital parameters listed above.

During the activity of XTE\,J1946+274 in 2010 and 2011, various X-ray
and gamma-ray missions observed the source (see \S\ref{sec:obs} and
Table~\ref{table:obs} for details), such that the pulse period
evolution is known in great detail especially from \textsl{Fermi}-GBM.
We searched for pulsations near the GBM period for
\textsl{Suzaku}-PIN, \textsl{RXTE}-PCA, and \textsl{Swift}-XRT using
the epoch folding technique. For the PIN we determined a pulse period
of 15.750025(27)\,s, see \S\ref{subsec:pulse}. For PCA, we used PCU2
top-layer light curves, extracted in GoodXenon mode with a time
resolution of 0.125\,s. The XRT data were taken in Windowed Timing
mode. The XRT light curves were obtained from a $\sim0\farcm5$ region
centered on the source position and rebinned to a 1\,s time
resolution. The initial uncertainties of the measured pulse periods
were estimated by Monte Carlo simulations, where synthetic light
curves of the source based on the observed pulse profile were searched
for the pulse period. The uncertainties of the periods measured by
\textsl{Fermi}-GBM were provided by the GBM Pulsar Project. The
measured pulse periods of XTE\,J1946+274 are shown in
Figure~\ref{fig:gbm_profile}.

In order to compute the pulse periods via
equation~(\ref{eq:periodtime}), we used the 1\,d binned, 15--50\,keV
\textsl{Swift}-BAT light curve of the source as the bolometric flux
evolution $F(t)$ and choose
$F_\mathrm{ref}=1\,\mathrm{count}\,\mathrm{s}^{-1}\,\mathrm{cm}^{-2}$.
Using the hard BAT flux as a proxy for the bolometric flux is
justified since the source does not show strong spectral changes over
and between outbursts \citep[][this work]{muller:12}. The main source
of uncertainty in the predicted pulse period therefore does not come
from changes in the spectral shape, but from the overall uncertainty
in the BAT flux measurements, which can have uncertainties of up to
15\%. In order to take these uncertainties into account, we use a
Monte Carlo approach in which 10000 BAT lightcurves are simulated. For
each time with a BAT measurement, $t_i$, we draw a simulated BAT count
rate from a Gaussian distribution with mean and standard deviation
given by the measured BAT rate and uncertainty, respectively. For each
of the light curve realizations we then derive the best-fit pulse
period evolution using equation~(\ref{eq:periodobs}). The standard
deviation of the resulting simulated pulse periods, $\sigma_{M(t_i)}$,
at each $t_i$ is then taken to be representative of the uncertainty of
the modeled pulse period evolution.

In order to obtain the final orbit and pulse period model, based on
an initial estimate for $\sigma_{M(t_i)}$ we minimize the fit statistics
\begin{equation}\label{eq:periodstat}
  \chi^2 = \sum_i \frac{(P_i - P_\mathrm{obs}(t_i))^2}{\sigma_{P_i}^2 + \sigma_{M(t_i)}^2}
\end{equation}
where $P_i$ is the measured pulse period at time $t_i$,
$P_\mathrm{obs}(t_i)$ is the model period
(equation~[\ref{eq:periodobs}]), and $\sigma_{P_i}$ and $\sigma_{M(t_i)}$
are the uncertainties of the data and the model as described above. We
then iteratively apply the Monte Carlo approach above to refine the
estimated model uncertainties. Usually three iterations are sufficient
to obtain convergence. Figure~\ref{fig:gbm_profile}d displays the
final estimate for the uncertainty of the pulse period model.

Fits to equation~(\ref{eq:periodobs}) are shown in
Figure~\ref{fig:gbm_profile}. The modeled intrinsic spin period $P(t)$
of the neutron star (shown in purple) dominates the period evolution
(black) compared to the effect of the orbital motion (orange). The two
residual panels show different assumptions for the exponent $\alpha$
of equation~(\ref{eq:pdotrelation}). In order to check the dependency
of the orbital parameters on the assumed torque model, we model the
data for both, $\alpha=6/7$ (the DISK model) and for $\alpha=1$ (the
WIND model). As illustrated by Figure~\ref{fig:gbm_profile}, both
models result in a successful description of the measured pulse period
evolution and yield \textsl{orbital} parameters that are consistent
with each other (Table~\ref{tab:periodfit}).

We stress again that for each of the two models the additional
uncertainties due to the BAT data have to be calculated separately by
the iterative Monte Carlo approach described above. The resulting
uncertainties of the model vary between 0.02 and 0.38\,ms with a mean
of 0.09\,ms (see Figure~\ref{fig:gbm_profile}d). Within the model
uncertainties, however, the model pulse periods agree with the
measured data. For example, the pulse period predicted by the DISK
model for the time of the \textsl{Suzaku} observation is
15.750300(380)\,s, while the observed period is 15.750025(27)\,s.
Unfortunately, the model uncertainty is large enough that it is not
possible for us to distinguish between the different torquing models,
with both model fits yielding almost the same $\chi^2$. Thankfully, as
shown in Table~\ref{tab:periodfit}, the orbital parameters are
insensitive to the details of modeling $\dot{P}(t)$. It is only the
best fit values for the spin change, $a$, and the torque strength,
$b$, which differ significantly. Numerical experimenting revealed that
this is due to a strong parameter degeneracy of the luminosity
exponent $\alpha$ with $a$ and $b$. Based on the pulse period
evolution alone it is therefore not possible to distinguish between
the two torquing scenarios.

\begin{deluxetable}{lcc} 
\tablecolumns{3} 
\tablewidth{\columnwidth} 
\tablecaption{Orbital parameters and spin period evolution. The
  uncertainties are on the 90\% confidence level.}
\startdata
\hline\hline
      & DISK & WIND \\
\hline
$a_\mathrm{sm} \sin i$ [lt-s]
      & $471.2^{+2.6}_{-4.3}$
      & $471.1^{+2.7}_{-2.8}$ \\
$P_\mathrm{orb}$ [d]
      & $172.7^{+0.6}_{-0.6}$
      & $171.4^{+0.4}_{-0.4}$ \\
$\tau$ [MJD]
      & $55514.8^{+0.8}_{-1.1}$
      & $55515.5^{+0.8}_{-0.7}$ \\
$e$
      & $0.246^{+0.009}_{-0.009}$
      & $0.266^{+0.007}_{-0.007}$ \\
$\omega$ [$^\circ$]
      & $-87.4^{+1.5}_{-1.7}$
      & $-87.1^{+1.2}_{-1.0}$ \\
$t_0$
      & $55550$ (fixed)
      & $55550$ (fixed) \\
$P_0$ [s]
      & $15.749742^{+0.000023}_{-0.000014}$
      & $15.749753^{+0.000013}_{-0.000013}$ \\
$a$ [s\,s$^{-1}$]
      & $1.67^{+0.16}_{-0.18} \times 10^{-10}$
      & $0.47^{+0.20}_{-0.10} \times 10^{-10}$ \\
$b$ [s\,s$^{-1}$]
      & $6.52^{+0.06}_{-0.08} \times 10^{-8}$
      & $10.76^{+0.05}_{-0.04} \times 10^{-8}$ \\
$\alpha$
      & $6/7$ (fixed)
      & $1$ (fixed) \\
${\chi}^{2}_{\mathrm{red}}/\mathrm{dof}$
      & 1.05/89
      & 1.06/89
\enddata 
\tablecomments{Listed are the projected semi major axis,
  $a_\mathrm{sm} \sin i$, the orbital period, $P_\mathrm{orb}$, the
  time of periastron passage, $\tau$, the eccentricity, $e$,
  the longitude of periastron, $\omega$, the reference time, $t_0$,
  the spin period at $t_0$, $P_0$, the constant spin-change, $a$, the
  torque strength, $b$, and the luminosity exponent, $\alpha$.}
\label{tab:periodfit}
\end{deluxetable} 

\section{SPECTRAL ANALYSIS}\label{sec:spec}

\begin{figure}
  \includegraphics[width=\columnwidth]{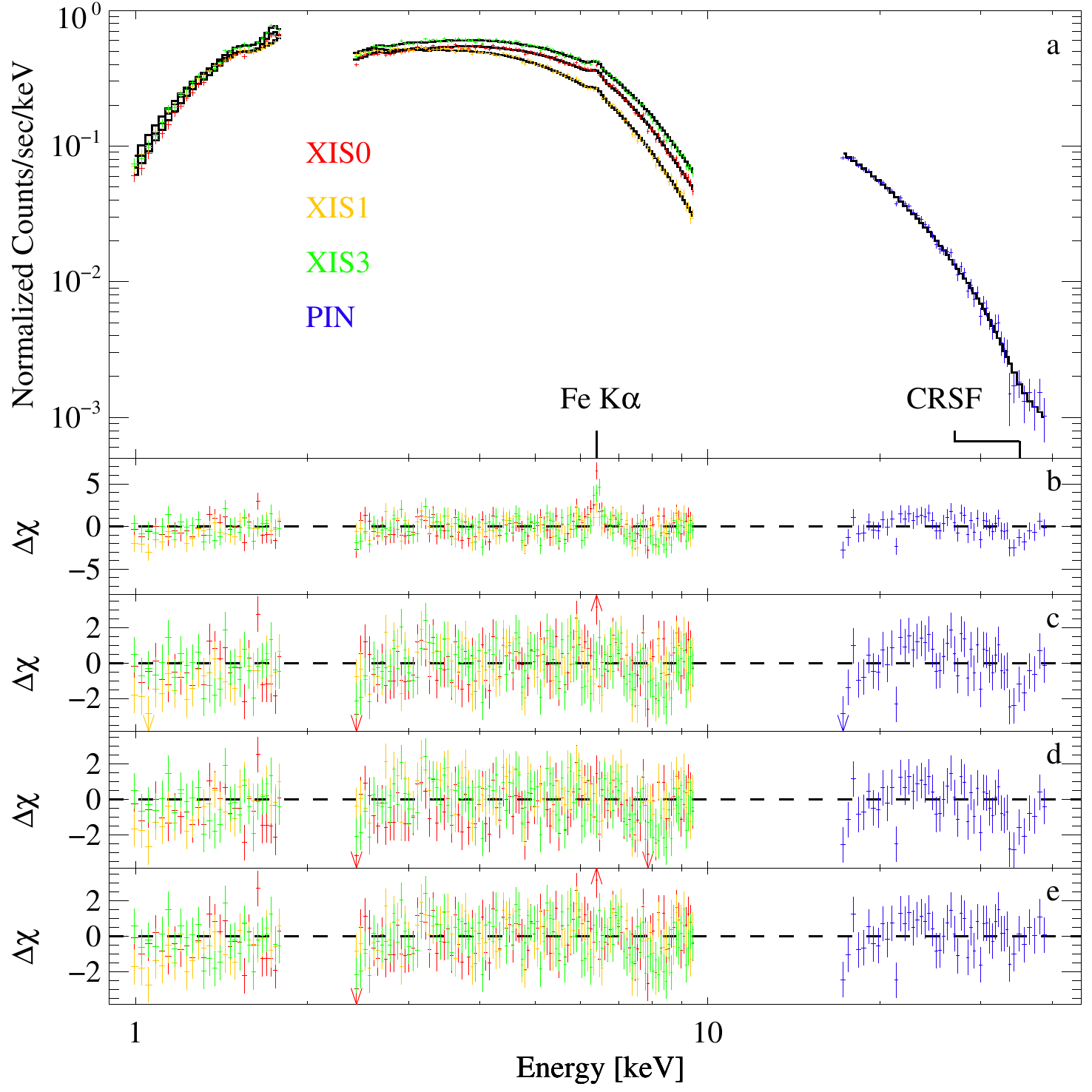}
  \caption{Spectra and best fit model for XIS 0, 1, 3 (in red, yellow,
    and green, respectively) and PIN (in blue). The spectra were
    fitted simultaneously with the model described by
    equation~(\ref{eq:model}) with an FDCO continuum model. The bottom
    panels show the residuals as $\Delta\chi$ obtained by (b) fitting
    only the continuum, (c) fitting the continuum with the
    Fe\,K$\alpha$ line, (d) fitting the continuum with the
    Fe\,K$\alpha$ line and the 35\,keV CRSF feature with
    $D_\mathrm{CRSF}$ set to 0 \textsl{after} fitting, (e) fitting the
    continuum with the Fe\,K$\alpha$ line and the 35\,keV CRSF
    feature.}\label{fig:spectra}
\end{figure}

\begin{deluxetable*}{ccccccc} 
\tablecolumns{7} 
\tablewidth{\textwidth} 
\tablecaption{Spectral fit parameters}
\tablehead{\colhead{ }  & \colhead{FDCUT\,I} & \colhead{FDCUT\,II} & \colhead{FDCUT\,III} & \colhead{CUTOFFPL} & \colhead{HIGHECUT} & \colhead{NPEX$^b$}}
\startdata

$N_\mathrm{H}\,[\times10^{22}\mathrm{cm}^{-2}]$ & $1.14^{+0.01}_{-0.02}$ & $1.66^{+0.02}_{-0.03}$ & $1.67(3)$ & $1.59^{+0.02}_{-0.04}$ & $1.59^{+0.02}_{-0.04}$ & $1.58^{+0.03}_{-0.05}$ \\

$A_\Gamma$[$\times 10^{-2}$\,keV$^{-1}$\,cm$^{-2}$\,s$^{-1}$] & $2.04^{+0.03}_{-0.05}$ & $2.02^{+0.03}_{-0.05}$ & $2.05^{+0.04}_{-0.05}$ & $0.97(2)$ & $0.97(2)$ & $0.96^{+0.02}_{-0.04}$ \\

$\Gamma$ & $0.55^{+0.01}_{-0.02}$ & $0.55^{+0.01}_{-0.02}$ & $0.57(2)$ & $0.41^{+0.02}_{-0.04}$ & $0.41^{+0.02}_{-0.04}$ & $0.39^{+0.04}_{-0.08}$ \\ 

$E_{\mathrm{fold}}$\,[keV] & $8.6^{+0.2}_{-0.3}$ & $8.5^{+0.2}_{-0.3}$ & $8.9^{+0.4}_{-0.4}$ & $9.6^{+0.4}_{-0.6}$ & $9.6^{+0.4}_{-0.6}$ & $9.1^{+0.8}_{-1.4}$ \\

$E_{\mathrm{cut}}$\,[$\times10^{-2}$\,keV] & $0.09^{+0.04}_{-0.09}$ & $0.01^{+0.00}_{-0.01}$ & $0.05^{+0.03}_{-0.05}$ & -- & $0.01^{+0.00}_{-0.01}$ & -- \\ 


$\Gamma_2$ & -- & -- & -- & -- & -- & -2$^a$ \\ 

$\alpha[\times10^{-2}]$\,[keV/keV] & -- & -- & -- & -- & -- & $0.020^{0.003}_{-0.020}$ \\ 

$E_{\mathrm{Fe}}$\,[keV] & -- & 6.41(3) & 6.41(3) & 6.41(3) & 6.41(3) & 6.41(3) \\ 

$\sigma_{\mathrm{Fe}}$\,[keV] & -- & 0.1$^a$ & 0.1$^a$ & 0.1$^a$ & 0.1$^a$ & 0.1$^a$ \\ 

$A_{\mathrm{Fe}}\,[\times 10^{-5}$\,photons\,cm$^{-2}$\,s$^{-1}$] & -- & $8.6^{+1.3}_{-1.4}$ & $8.6^{+1.3}_{-1.3}$ & $8.7^{+1.3}_{-1.4}$ & $8.7^{+1.3}_{-1.4}$ & $8.7^{+1.1}_{-1.0}$ \\ 

$E_{\mathrm{CRSF}}$\,[keV] & -- & -- & $35.2^{+1.5}_{-1.3}$ & $34.8^{+1.2}_{-1.0}$ & $34.8^{+1.2}_{-1.0}$ & $34.8^{+1.1}_{-1.0}$ \\ 

$\sigma_{\mathrm{CRSF}}$\,[keV] & -- & -- & 2$^a$ & 2$^a$ & 2$^a$ & 2$^a$ \\ 

$D_{\mathrm{CRSF}}$\,[keV] & -- & -- & $2.4^{+1.5}_{-1.3}$ & $3.5^{+1.5}_{-1.5}$ & $3.5^{+1.5}_{-1.5}$ & $3.8^{+1.6}_{-1.5}$ \\ 

$c_{\mathrm{XIS\,0}}$ & 1$^a$ & 1$^a$ & 1$^a$ & 1$^a$ & 1$^a$ & 1$^a$ \\

$c_{\mathrm{XIS\,1}}$ & $1.07(1)$ & $1.07(1)$ & $1.07(1)$ & $1.07(1)$ & $1.07(1)$ & $1.07(1)$ \\

$c_{\mathrm{XIS\,3}}$ & $0.95(1)$ & $0.95(1)$ & $0.95(1)$ & $0.95(1)$ & $0.95(1)$ & $0.95(1)$ \\

$c_{\mathrm{PIN}}$ & $1.32^{+0.07}_{-0.05}$ & $1.35^{+0.07}_{-0.05}$ & $1.29^{+0.07}_{-0.06}$ & $1.29^{+0.07}_{-0.05}$ & $1.29^{+0.08}_{-0.05}$ & $1.31^{+0.07}_{-0.06}$ \\

${\chi}^{2}_{\mathrm{red}}/\mathrm{dof}$ & $1.38/470$ & $1.19/468$ & $1.17/466$ & $1.12/467$ & $1.12/466$ & $1.12/466$

\enddata 
\tablecomments{The XIS and PIN spectra were fitted simultaneously with
  the models described in \S\ref{subsec:model}. The columns are
  labeled according to the continuum that was used. The uncertainties
  are given on a 90\% confidence level.  $^a$ These parameters were
  frozen while fitting; $^b$ for the model $M_\mathrm{NPEX}(E) \propto
  (E^{-\Gamma} + \alpha E^{+\Gamma_2})e^{-E/E_\mathrm{fold}}$ the
  parameters $\Gamma$ and $\Gamma_2$ are the indices of the falling
  and rising power law components and $\alpha$ is the normalization of
  the rising relative to the falling component.}
\label{table:spec_params1}
\end{deluxetable*} 

\subsection{Best Fit Model}\label{subsec:model}

We modeled the 1--9.4\,keV XIS and the 17--38\,keV PIN spectra using
\texttt{xspec12} \citep{arnaud:96}. The 1.8--2.4\,keV range was
excluded due to known calibration uncertainties \citep{abc_guide}. We
applied the normalization constants $c_{\mathrm{XIS\,1}}$,
$c_{\mathrm{XIS\,3}}$, and $c_{\mathrm{PIN}}$ to account for the flux
cross-calibration between the respective instruments relative to
XIS\,0, where $c_{\mathrm{XIS\,0}}$ was fixed at 1 (\texttt{xspec}
model \texttt{constant}). The absorption was modeled with
\texttt{tbnew}, an updated version of
\texttt{tbabs}\footnote{\url{http://pulsar.sternwarte.uni-erlangen.de/wilms/research/tbabs/}},
using cross sections by \citet{verner:95} and abundances by
\citet{wilms:00}. Extending the fit down to 0.8\,keV,
\citet{maitra:13} included an additional partial covering absorption
component. Since they found that its parameters are model dependent
and since the hardness ratio evolution over the observation
(Figure~\ref{fig:lc_hardness}c and Figure~\ref{fig:lc_hardness}d) does
not indicate any variability due to partial covering, we used one
fully covering absorber alone which is suffient to model the data down
to 1\,keV well.

Following the spectral analysis of \citet{muller:12}, we first fitted
a Fermi-Dirac cutoff model
\citep[\texttt{power$\times$fdcut},][]{tanaka:86}, described by:
\begin{equation}
M_\mathrm{FDCUT}(E)\propto E^{-\Gamma}\times\left[1+\exp\left(\frac{E-E_\mathrm{cut}}{E_\mathrm{fold}}\right)\right]^{-1}
\end{equation}
where the photon flux at energy $E$ is described by a power law with a
photon index $\Gamma$, multiplied by an exponential cutoff at energy
$E_{\mathrm{cut}}$ with a folding energy $E_{\mathrm{fold}}$. The soft
Galactic ridge emission seen in the 6--7\,keV range, which needed to
be taken into account for PCA data modeling by \citet{muller:12}, is
not required for \textsl{Suzaku} due to XIS being an imaging
instrument. The results of this fit are listed in
Table~\ref{table:spec_params1} in the column labeled FDCUT\,I.
Figure~\ref{fig:spectra}b shows the residuals from fitting the
continuum model only.

The strongest residuals are seen at 6.41\,keV. We interpreted this as
a narrow Fe\,K$\alpha$ fluorescence line that we proceeded to describe
with a Gaussian line model (\texttt{gaussian}). The width is
unresolved and we fixed it at $\sigma_{\mathrm{Fe}}=0.1$\,keV,
slightly below the XIS detector resolution. The results of this fit
are listed in Table~\ref{table:spec_params1} in the column labeled
FDCUT\,II and Figure~\ref{fig:spectra}c shows the fit residuals.

Residuals are still visible in the PIN energy range, especially around
35\,keV. We included an absorption-like line with a Gaussian optical
depth profile (\texttt{gabs}) often used to describe cyclotron lines:
\begin{equation}
M_\mathrm{CRSF}(E)=\mathrm{exp}(-\tau(E))
\end{equation}
with
\begin{equation}
\tau(E)=\tau_{\mathrm{CRSF}}\mathrm{exp}\left[-\frac{1}{2}\left(\frac{E-E_{\mathrm{CRSF}}}{\sigma_{\mathrm{CRSF}}}\right)^{2}\right]
\end{equation}
where $E_\mathrm{CRSF}$ is the cyclotron line energy,
$\sigma_\mathrm{CRSF}$ is the line width, and $\tau_\mathrm{CRSF}$ is
the optical depth. Note that the \texttt{gabs} implementation provides
the line depth $D_\mathrm{CRSF} = \tau_\mathrm{CRSF} 
\sigma_\mathrm{CRSF} \sqrt{2\pi}$ instead of
$\tau_\mathrm{CRSF}$. The CRSF width was unresolved and we fixed it at
$\sigma_{\mathrm{CRSF}}=2$\,keV, close to PIN's detector resolution.
The results of this fit are listed in Table~\ref{table:spec_params1}
in the column labeled FDCUT\,III. Figure~\ref{fig:spectra}a shows the
spectra and fitted model and Figure~\ref{fig:spectra}e shows the fit
residuals. The latter do not show any further strong features. In
order to illustrate the contribution of the CRSF feature to the best
fit Figure~\ref{fig:spectra}d shows the residuals of the best fit with
the CRSF depth set to 0. We tried fixing the PIN cross normalization
constant to its canonical value of 1.181 for an HXD-nominal pointing
position \citep{suzaku_memo2008_06}. This resulted in a worse fit with
$\chi^2_\mathrm{red}=1.35$, therefore, we left $c_\mathrm{PIN}$ free.

We then checked whether the presence of a ``10\,keV feature'' is
consistent with the data. This is a broad residual that has been
observed in the spectra of several accreting pulsars thought to be
caused by imperfect modeling of the continuum shape using empirical
models \citep[see, e.g.,][]{coburn:02}. It is generally detected as a
positive residual \citep[e.g., in Cen~X-3, see][]{suchy:08} but in
some sources, including XTE\,J1946+274, it appears as a negative one
\citep[e.g., in \mbox{Vela~X-1}, see][]{furst:14}. We applied the
deeper of the two detections reported for XTE\,J1946+274 by
\citet{muller:12} to our model, i.e., following them we included a
\texttt{gauabs} component (another parametrization of the
\texttt{gabs} shape) with $E_\mathrm{10\,keV}=9.85$\,keV,
$\sigma_\mathrm{10\,keV}=2.2$\,keV, and $\tau_\mathrm{10\,keV}=0.069$.
This approach did not significantly change the quality of the fit and
fitting $\tau_\mathrm{10\,keV}$ resulted in a value consistent with 0.
We conclude that such a component could be present in the spectrum but
is not detected, probably in part due to the lack of data between 9.4
and 17\,keV.

Our \texttt{fdcut} based best fit model (FDCUT\,III) thus consists of
absorption in the interstellar medium as well as intrinsic to the
system, a power law continuum with a rollover, a Gaussian emission
line for Fe\,K$\alpha$ fluorescence, and an absorption-like line with
a Gaussian optical depth profile for the cyclotron line:
\begin{equation}
M_\mathrm{best}(E) = \texttt{const} \times \texttt{tbnew} \times (\texttt{power} \times \texttt{fdcut} + \texttt{gauss}) \times \texttt{gabs}
\label{eq:model}
\end{equation}
in \texttt{xspec} notation. We obtain an unabsorbed 3--60\,keV flux of
$4.40\pm0.01 \times 10^{-10}
\,\mathrm{erg}\,\mathrm{s}^{-1}\,\mathrm{cm}^{-2}$.

In the following we present results replacing the
\texttt{power$\times$fdcut} continuum with other continuum models
commonly applied to accreting X-ray pulsars \citep[see, e.g.,][for the
  equations describing these models]{muller:13}: a power law with an
exponential cutoff (\texttt{cutoffpl}), a power law with a high energy
cutoff (\texttt{power$\times$highecut}, sometimes also called
\texttt{plcut}), and the sum of a negative and a positive power law
with an exponential cutoff \citep[\texttt{npex},][]{mihara:95}. The
last three columns of Table~\ref{table:spec_params1} show the best fit
spectral parameters using these continuum models. Since the fitted
values of the cutoff energy of \texttt{highecut} and the normalization
of the positive power law of \texttt{npex} are consistent with 0 these
three models are degenerate and result in the same fit quality and in
the same values of their common parameters. The \texttt{fdcut} fit has
a slightly different rollover shape but its parameters are also
qualitatively, and often quantitatively within errors, the same. We
note that the \texttt{npex} parameters reported by \citet{heindl:01}
for the bright outburst of 1998, which were obtained fitting averaged
\textsl{RXTE} monitoring spectra above 8\,keV can also describe the
PIN spectrum, but they do not provide a good description of the XIS
spectrum (below 8\,keV the spectra were variable between individual
monitoring pointings). 

\citet{maitra:13} reported \texttt{highecut} and \texttt{npex} fit of
the same \textsl{Suzaku} dataset. Their best fit parameters are
generally not consistent with ours. For example, their
\texttt{highecut} cutoff energy of $7.02^{+0.69}_{-0.29}$\,keV and
their \texttt{npex} positive power law normalization are not
consistent with 0. A possible explanation for this discrepancy is that
the \texttt{highecut} model has a break at the cutoff energy, which
here is located at the energy of the Fe K edge. In part this approach
therefore could be modeling imperfections of the fit in the region of
the iron line and edge. No edge component was required in our fits.
Using the approach of \citet{maitra:13} by extending the spectrum to
70\,keV, i.e., beyond where the source is detected (see next section),
and allowing for a 9\,keV wide cyclotron line using the
\texttt{cyclabs} model, did, we were able to reproduce their continuum
parameters. \citet{maitra:13} do not quote flux calibration constants.
We found a PIN/XIS ratio similar to our other fits. We also assumed
that their unitless CRSF width values $W_\mathrm{CRSF}$ were given in
keV. As mentioned in \S\ref{sec:intro} a 9\,keV wide cyclotron line
can be expected to in part model the continuum \citep{muller:13}.
Similar to \citet{maitra:13} we found that thermal Comptonization of
soft photons in a hot plasma \citep[\texttt{comptt},][]{titarchuck:94}
cannot explain the \textsl{Suzaku} spectra, particularly in the PIN
range ($\chi^2_\mathrm{red}$/dof=5.86/464, unconstrained parameters).

\subsection{Cyclotron Resonance Scattering Feature}\label{subsec:crsf}

\begin{figure}
  \includegraphics[width=\columnwidth]{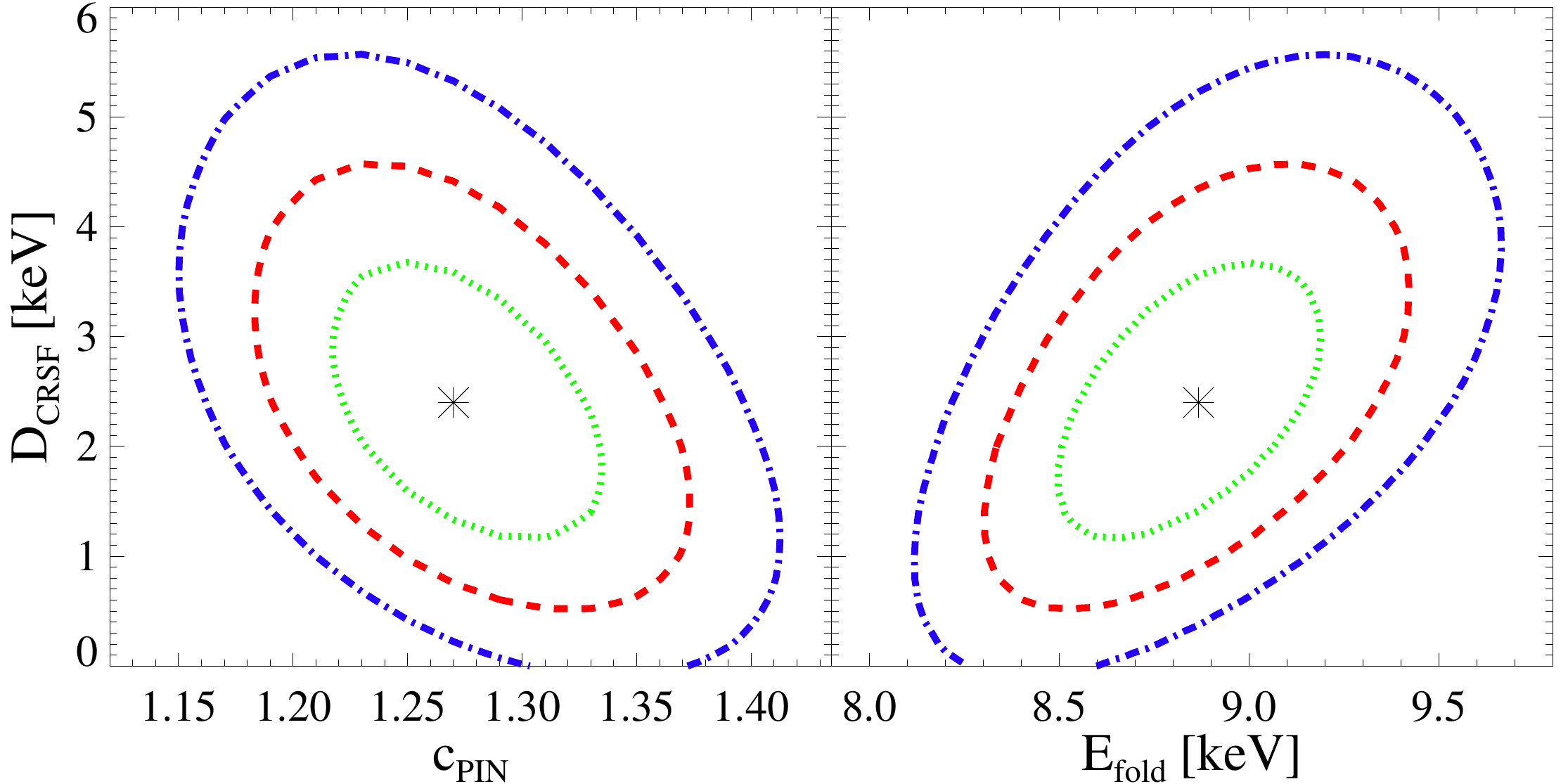}
  \caption{Confidence contour plots showing moderate correlations
    between the depth of the CRSF at 35\,keV and the folding energy
    (top) and between the depth of the CRSF at 35\,keV and the PIN
    flux cross-calibration constant (bottom) for the FDCUT\,III fit of
    Table~\ref{table:spec_params1}. Contours for confidence levels of
    1, 2, and 3$\sigma$ are shown in dotted green, dashed red, and
    dash-dotted blue, respectively.}\label{fig:contours}
\end{figure}

The cyclotron line we found in the \textsl{Suzaku} spectrum from the
end of the second outburst of the 2010 series has an energy of
$E_{\mathrm{CRSF}}=35.16^{+1.5}_{-1.3}$\,keV, a line depth of
$D_{\mathrm{CRSF}}=2.42^{+1.5}_{-1.3}$\,keV and a fixed width of
$\sigma_{\mathrm{CRSF}}=2\,$keV. \citet{heindl:01} found a CRSF with
similar parameters at $E_{\mathrm{CRSF}}=36.2^{+0.5}_{-0.7}$\,keV with
$D_{\mathrm{CRSF}}=2.79^{+2.14}_{-1.77}$\,keV ($\tau_{\mathrm{CRSF}}=
0.33^{+0.07}_{-0.06}$) and
$\sigma_{\mathrm{CRSF}}=3.37^{+0.92}_{-0.75}$\,keV) for the bright
outburst in 1998, from \textsl{RXTE} data. The CRSF energy obtained
with \textsl{RXTE} is consistent with the one obtained with
\textsl{Suzaku}. \citet{muller:12} did not find a line at 35\,keV, but
found marginal evidence (1.8$\sigma$) for a CRSF at $\sim$25\,keV in
the first and third outbursts of the 2010 series.

In order to check the robustness of the \textsl{Suzaku} detection of
an unresolved cyclotron line at 35\,keV with resepct to changes of the
continuum model parameters we calculated confidence contours for two
parameters of interest, the CRSF depth $D_\mathrm{CRSF}$ and one
continuum parameter at a time. We found no strong correlations. Not
unexpectedly, moderate correlations are present with the folding
energy $E_\mathrm{fold}$ and with the flux cross-calibration constant
of the PIN spectrum $c_\mathrm{PIN}$, see Figure~\ref{fig:contours}.
The confidence contours indicate that the CRSF feature is present
independently of the continuum modeling on a $\sim3\sigma$ level. We
further confirmed this picture by determining a significance of
$2.81\sigma$ for a cyclotron line feature at 35\,keV using Monte Carlo
simulations. This significance value was obtained by simulating 5000
spectra based on the best fit model parameters without the CRSF
(column FDCUT\,II of Table~\ref{table:spec_params1}) and fitting them
with and without including the CRSF (width fixed at 2\,keV in the
former case). In 25 cases we found a bigger improvement in $\chi^2$
than in the real data, resulting in the quoted significance. For an
unresolved line at 25\,keV line we determined a $3\sigma$ upper limit
of $D_\mathrm{CRSF}\sim0.9$ for the line depth, based on Monte Carlo
simulations including a 25\,keV line with different depths and for
each depth comparing the $\chi^2$ values obtained from fitting the
line to the data and the simulations. 

We also investigated the modeled PIN background spectrum and the
effect of its uncertainty on the fit parameters, particularly of the
cyclotron line. To this end we first included the background
normalization as a fit parameter in the FDCUT\,III model using
\texttt{recorn}. The uncertainty of the fitted background
normalization ranged from a decrease of 20\% to an increase of 3\%.
Repeating the fit fixing the background normalization at either of
these values or at the default and adding the expected systematic
uncertainty of 3\% to the PIN background spectrum
\citep[][node10]{abc_guide} did not significantly change the resulting
cyclotron line parameters. We confirm \citet{maitra:13}'s report that
the normalization of the background spectrum observed when the source
was occulted by the Earth (obtained by setting $\texttt{ELV}<-5^\circ$
in \texttt{aepipeline}) was about 20\% below that of the modeled
background spectrum. This result can qualitatively be explained with
the anticorrelation between the magnetic field strength and the
background flux at a given satellite
location \citep[][node12]{tech_description}: A measure for the strength
of the Earth's magnetic field -- the time resolved magnetic cutoff
rigidity of the Earth at the satellite position during the observation
-- can be obtained from the observation's filter file and we found
that it was on average lower during the on-source time
($\texttt{ELV}>5$) than during the Earth-occultation time
($\texttt{ELV}<-5$) for the XTE\,J1946+274 observation.

In \S\ref{subsec:pulse} we showed that there is no broad-band
detection of the pulsar above 40\,keV. The background subtracted
spectrum generally confirms this. It is consistent with 0 above
38\,keV with the exception of two independent spectral bins in the
43--47\,keV range that show a marginal source detection \citep[see
also Figure~4 of][]{maitra:13}. The picture stays the same when taking
the 3\% background uncertainty into account. Using non background
subtracted events we detected no pulsations in the 38--45\,keV range
and marginal ones in the 43--47\,keV range, confirming again that the
background model is sufficiently accurate.  The background spectrum
dominates over the source contribution above $\sim$33\,keV and
declines smoothly with energy with no systematic features around 35 or
40\,keV. Above 38\,keV the source spectrum might thus show some
structure but it is mostly below the detection limit and was therefore
excluded from our analysis.

\section{DISCUSSION}\label{sec:discussion}

\subsection{Pulse Period Evolution and Orbit Parameters}

We successfully applied the accretion torque theory of
\citet{ghosh:79b} to XTE\,J1946+274 and updated the orbital solution
for this source (Table~\ref{tab:periodfit}). Previously
\citet{wilson:03} used three different approaches to model the
observed pulse period evolution obtained by \textsl{RXTE}-PCA and
\textsl{CGRO}-BATSE in 1998, which was dominated by a strong spin-up
as well. Comparing the resulting orbital parameters to ours we find
that the semi-major axis, $a_\mathrm{sm} \sin i$, agrees best
with their 10th-order polynomial model. Extrapolating our derived time
of periastron passage, $\tau$, back to 1998, gives times which agree
to within $2\sigma$ with the result of their model as well. The
orbital period, $P_\mathrm{orb}$, and eccentricity, $e$, are
consistent with their linear model, while the longitude of periastron,
$\omega$, is the same as in their piecewise approximation within the
uncertainties. As noted by \citeauthor{wilson:03}, however, the
$\chi^2$ of all three different approaches is not acceptable because
the models do ``not completely describe the intrinsic torques''. 

In contrast to other methods such as, e.g., a Fourier series approach
\citep[e.g.,][]{kuehnel:13}, calculating the spin-up of accreting
pulsars using the theory of \citet{ghosh:79b} allows us to model the
possibly complex, intrinsic spin period evolution of the neutron star
with better accuracy \citep[see also][]{galloway:04,sugizaki:15}. As a
result the orbital motion can be properly disentangled from the
overall observed pulse period evolution and the derived orbital
parameters are generally more reliable. We caution, however, that
assuming $P(t')$ is a constant on the right side of
equation~(\ref{eq:periodtime}) in order to simplify the calculation of
this differential equation \citep[see, e.g.,][]{sugizaki:15} might
lead to additional uncertainties when fitting longer time series. If
we set $P(t')=P_0$, for example, the modeled pulse period evolution
differs up to 0.01\,ms, which is of the same order as the
uncertainties of the \textsl{Fermi}-GBM period measurements. As soon
as more precise flux measurements are used for $F(t)$ or the measured
spin-up is even stronger than for XTE\,J1946+274 the differential
equation should thus be solved properly.

This kind of timing analysis would not be possible without regular
flux monitoring by all-sky observatories, such as \textsl{Fermi}-GBM,
\textsl{Swift}-BAT, or MAXI.

\subsection{Mass Function and Orbit Inclination}

The accurately determined orbital parameters allow us to derive the
value of the mass function of XTE\,J1946$+$274 following the same
approach as in, e.g., \citet{wilson:03}. The mass function of a
binary,
\begin{equation}\label{eq:massfunction}
  f(M) = \frac{(M_\mathrm{opt} \sin i)^3}{(M_\mathrm{NS} +
  M_\mathrm{opt})^2} = \frac{4 \pi^2}{G} \frac{(a_\mathrm{sm}
  \sin i)^3}{P_\mathrm{orb}^2}
\end{equation}
depends on the masses, $M_\mathrm{opt}$ and $M_\mathrm{NS}$, of the
optical companion and neutron star, respectively, and on the orbital
inclination angle, $i$. However, the mass function can also be
calculated using the orbital period, $P_\mathrm{orb}$, and the
projected semi-major axis, $a_\mathrm{sm} \sin i$. Using the orbital
parameters listed in Table~\ref{tab:periodfit} we derive consistent
values of $f(M)=3.77^{+0.11}_{-0.07}\,\mathrm{M}_\odot$ for
disk-accretion and $f(M)=3.82^{+0.07}_{-0.07}\,\mathrm{M}_\odot$ for
wind-accretion.  Assuming the same mass range for the companion star
of $10~\mathrm{M}_\odot \le M_\mathrm{opt} \le 16~\mathrm{M}_\odot$ as
used by \citet{wilson:03} and the canonical neutron star mass
$M_\mathrm{NS} = 1.4\,\mathrm{M}_\odot$, we can solve
equation~(\ref{eq:massfunction}) for the inclination angle, $i$. Using
the widest possible range for the mass function as calculated above,
$3.70~\mathrm{M}_\odot \le f(M) \le 3.89~\mathrm{M}_\odot$, we derive
an orbital inclination angle of $41^\circ \le i \le 52^\circ$. This is
in good agreement with the value of $i \gtrsim 46^\circ$ as found by
\citet{wilson:03}.

As already argued by \citet{wilson:03}, the inclination angle of the
Be-disk, $i_\mathrm{disk}$, with respect to the observer is not
necessarily aligned with the inclination angle of the orbit, $i$.
From measurements of the width of the single-peaked H$\alpha$ line in
an optical spectrum, \citet{wilson:03} concluded that the Be-star is
seen nearly pole-on. Thus, the Be-disk and orbital plane might indeed
be misaligned in XTE\,J1946+274. \citet{arabaci:14} recently analyzed
optical spectra of the system and noted, however, that deriving the
Be-disk inclination from the H$\alpha$ line profile is highly
uncertain based on theoretical investigations by
\citet{silaj:10}. Assuming that the orbital plane and the Be-disk are
aligned ($i_\mathrm{disk} = i$), \citet{arabaci:14} derived the
rotational velocity of the Be-star. They concluded that the Be
companion of XTE\,J1946+274 is rotating with 0.50--0.72 times the
critical break-up velocity of a typical Be type star
($v_\mathrm{crit}{\sim}618\,\mathrm{km}\,\mathrm{s}^{-1}$). Using
their initial value of the projected velocity, $v \sin i =
323\,\mathrm{km}\,\mathrm{s}^{-1}$ and our determined inclination
angle, $i$, we find a velocity of 0.66--0.80 times the break-up
velocity.

\subsection{Outburst Behavior}

Two outburst series of XTE\,J1946+274 have been observed, one in 1998
\citep{wilson:03} and one in 2010 (Figure\ref{fig:batlc}) with two to
three outbursts per orbit. In order to explain this X-ray activity the
companion of XTE\,J1946+274 has been studied in the optical and IR.
Based on observations of permanent H$\alpha$ emission,
\citet{arabaci:14} conclude that during X-ray quiescence a large Be
disk is present. They observed a brightening in the optical/IR
indicating that the Be star experienced a long mass-ejection event
from 2006 to 2012, reaching its maximum intensity in 2010, around the
time of the outburst series. \citeauthor{arabaci:14} postulate that
this ejection caused an increase in size, perturbations, and warping
of the Be disk. They also state that the X-ray activity is triggered
by the neutron star coming into contact with the warped areas in the
tilted Be disk. This could explain why we observe two to three
outbursts per orbit. The presence of H$\alpha$ and optical/IR
emissions after the X-ray activity indicates that once the material
was consumed through accretion, the Be disk quickly and steadily
recovered and the system returned to quiescence \citep{arabaci:14}.

\subsection{Continuum and Fe\,K$\alpha$ Line}\label{discussion:continuum}

We described the spectral shape of XTE\,J1946+274 with a Fermi Dirac
Cutoff power law together with an Fe K$\alpha$ fluorescence line and a
CRSF at 35\,keV. We find $E_\mathrm{fold}=8.89(4)\,$keV and a hydrogen
column density of $N_\mathrm{H}=1.67(3)\times10^{22}\mathrm{cm}^{-2}$.
These parameters are roughly consistent with the ones found by
\citet{muller:12} in PCA data taken during earlier outbursts, namely
$E_\mathrm{fold}=6.0^{+2.6}_{-1.6}$--$8.1^{+0.7}_{-0.6}$\,keV, and
$N_\mathrm{H}=1.77^{+0.25}_{-0.29}$--$5.1^{+2.5}_{-3.3}\times10^{22}\mathrm{cm}^{-2}$.
Their measured $\Gamma=0.74^{+0.12}_{-0.17}$--$1.04^{+0.13}_{-0.18}$
is slightly softer than ours, $\Gamma=0.57(2)$. The cutoff energy is
different as well: It is found here to be zero, while
\citet{muller:12} found $E_\mathrm{cut}=14\pm
4$--19.4$^{+2.1}_{-9.7}$\,keV.

In order to study the changes in the spectral shape at different times
and luminosities during the outburst series we compared our best fit
model and the models fitted in \citet{muller:12} by eye. The
\textsl{Suzaku} spectrum is harder at high energies ($>$12\,keV) than
the spectra from \citet{muller:12}. This hardness change could be an
indication of a higher temperature of the plasma in the accretion
column, despite the lower luminosity. At first glance this may seem
inconsistent, however, the electron temperature and mass accretion
rate cannot be clearly determined without a physical continuum model.
The implementation and testing of such a physical model is work in
progress \citep{marcu:14}.

\begin{figure}
  \includegraphics[width=\columnwidth]{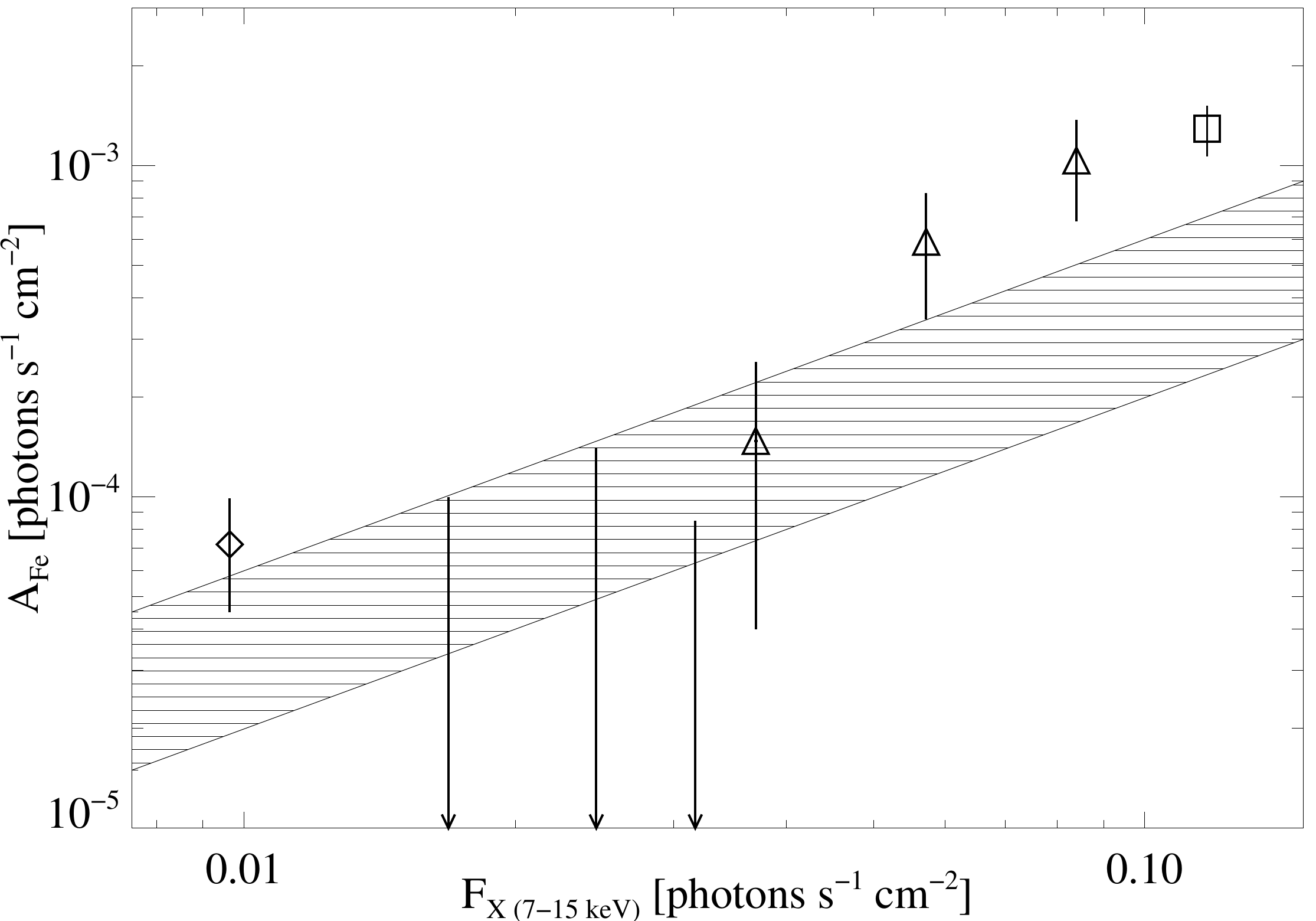}
  \caption{Flux of the Fe\,K$\alpha$ fluorescence line as a function
    of the 7--15\,keV continuum flux. The diamond represents the
    \textsl{Suzaku}-XIS\,0 data. All other data points are from
    \citet{muller:12}: triangles represent multiple instrument results
    from the first 2010 outburst (\textsl{RXTE, INTEGRAL} -- two
    high-flux triangles) and from the third 2010 outburst
    (\textsl{Swift, RXTE, INTEGRAL} -- two low-flux triangles and two
    upper limits), and the square corresponds to the \textsl{RXTE}
    average spectrum of the 1998 outburst. The hashed region describes
    the predicted correlation according to \citet{nagase:86}
    calculated using $N_\mathrm{H}$ and continuum normalization values
    from the \textsl{Suzaku} spectral fit and the \citet{muller:12}
    fits.}\label{fig:fe_flux}
\end{figure}

Both, \citet{muller:12} and we find $N_\mathrm{H}$ values that are
almost twice as large as the Galactic $N_\mathrm{H}$ in the direction
of XTE\,J1946+274
\citep[$N_\mathrm{H}=9.4\times10^{21}\mathrm{cm}^{-2}$;][]{kalberla:05}.
This excess indicates the presence of absorbing material intrinsic to
the X-ray binary system. The excitation of such neutral to moderately
ionized material surrounding the neutron star by the X-rays emitted
from the accretion column can produce fluorescent lines from iron and
other elements. These lines are a very useful tool for analyzing the
properties of the material
\citep[e.g.,][]{inoue:85,leahy:93,torrejon:10,reig:13}.
 
We find a narrow ($\sigma_\mathrm{Fe}=0.1$\,keV) Fe\,K$\alpha$
fluorescent emission line at $E_\mathrm{Fe}=6.41(3)\,$keV, confirming
the presence of this neutral to moderately ionized material. The flux
was $A_{\mathrm{Fe}}\sim 8.6\times
10^{-5}\,\mathrm{photons}\,\mathrm{cm}^{-2}\,\mathrm{s}^{-1}$ (see
Table~\ref{table:spec_params1}). The equivalent width is 32.2\,eV for
the \textsl{Suzaku} observation, consistent with the $\sim$29\,eV
found by \citet{maitra:13} in the same data set, but lower than the
measured 49--69\,eV found in earlier data taken at different fluxes
\citep{heindl:01,muller:12}. As shown, e.g., by \citet{inoue:85}, one
expects the flux in the flourescence line to be correlated with the
continuum flux above 7\,keV. Figure~\ref{fig:fe_flux} shows this
relationship using data of all published observations of
XTE\,J1946+274, extending a similar figure by \citet{muller:12} to
lower fluxes. The figure also shows the correlation predicted by
equation~(4a) of \citet{nagase:86} which is an estimate for the
fluorescent line flux as a function of $N_\mathrm{H}$ and continuum
flux. The hashed region in Figure~\ref{fig:fe_flux} illustrates the
range of the expected Fe\,K$\alpha$ flux values according to
\citet{nagase:86}, taking into account the variation in $N_\mathrm{H}$
between all published spectral fits. This range is an upper limit to
the absorption column of the system. For the values with the lowest
uncertainties the observed Fe\,K$\alpha$ flux is slightly higher than
the one predicted by \citet{nagase:86}. This is especially the case
for the high flux data points and is qualitatively consistent with
their higher equivalent width compared to the \textsl{Suzaku}
measurement. A possible reason for this slight excess could be an
overabundance of iron in the emitting medium. Alternatively, the
excess could also be due to the fact that the ionization structure of
the material is more complicated than the purely neutral Fe absorber
assumed by \citet{nagase:86}. Finally, it is also likely that the
emission is not purely from the line of sight, but from other areas
such as fluorescence from a tilted and/or warped Be disk around the
neutron star.

\subsection{Cyclotron Resonance Scattering Feature}\label{discussion:crsf}

Evidence of a $\sim$35\,keV cyclotron line line was first seen by
\citet{heindl:01} in \textsl{RXTE} data obtained during a time when
the source was much brighter than in the observations analyzed here.
Our \textsl{Suzaku} observation supports the presence of this line:
the $\chi^2$ slightly improved from 557 to 545, corresponding to a
significance of $2.81\sigma$ (obtained using Monte Carlo simulations),
between the FDCUT\,II and FDCUT\,III fits. Including this line
improved the fits with the other continuum models as well. The
centroid energy of $35.2^{+1.5}_{-1.3}$\,keV implies a surface
magnetic field of $B_\mathrm{NS} = 3.1^{+0.1}_{-0.1}(1+z)\times
10^{12}$\,G.

The CRSF parameters are independent of the continuum model.
Furthermore, describing the PIN data only with the \texttt{npex}
model, we obtain a good fit with ${\chi}^{2}_{\mathrm{red}}=0.92$ for
50 d.o.f., for continuum parameters consistent with
\citet{heindl:01}. Both the energy of the cyclotron line and its
optical depth measured with \textsl{Suzaku} are within $1\sigma$ of
those measured with \textsl{RXTE}. Note, however, that due to spectral
complexity below 10\,keV the \textsl{RXTE} based \texttt{npex} values
do not describe the broad band (XIS and PIN) \textsl{Suzaku} data.

We find a lower centroid energy for the CRSF than the effectively
$\sim$40\,keV previously reported for this dataset by
\citet{maitra:13}. As explained in \S\ref{subsec:crsf}, their higher
value could be due to in part modeling an artificial feature, as these
authors include PIN data above 40\,keV, where the source is mostly not
detected.

Our spectrum is not consistent with the 25\,keV feature discussed by
\citet{muller:12}. We tried including a feature with their parameters
and the ${\chi}^2_\mathrm{red}$ increased to 1.92. When the depth of
this 25\,keV feature was left free it became consistent with zero.

\subsection{Accretion Column}\label{discussion:acc_column}

\begin{figure}
  \includegraphics[width=\columnwidth]{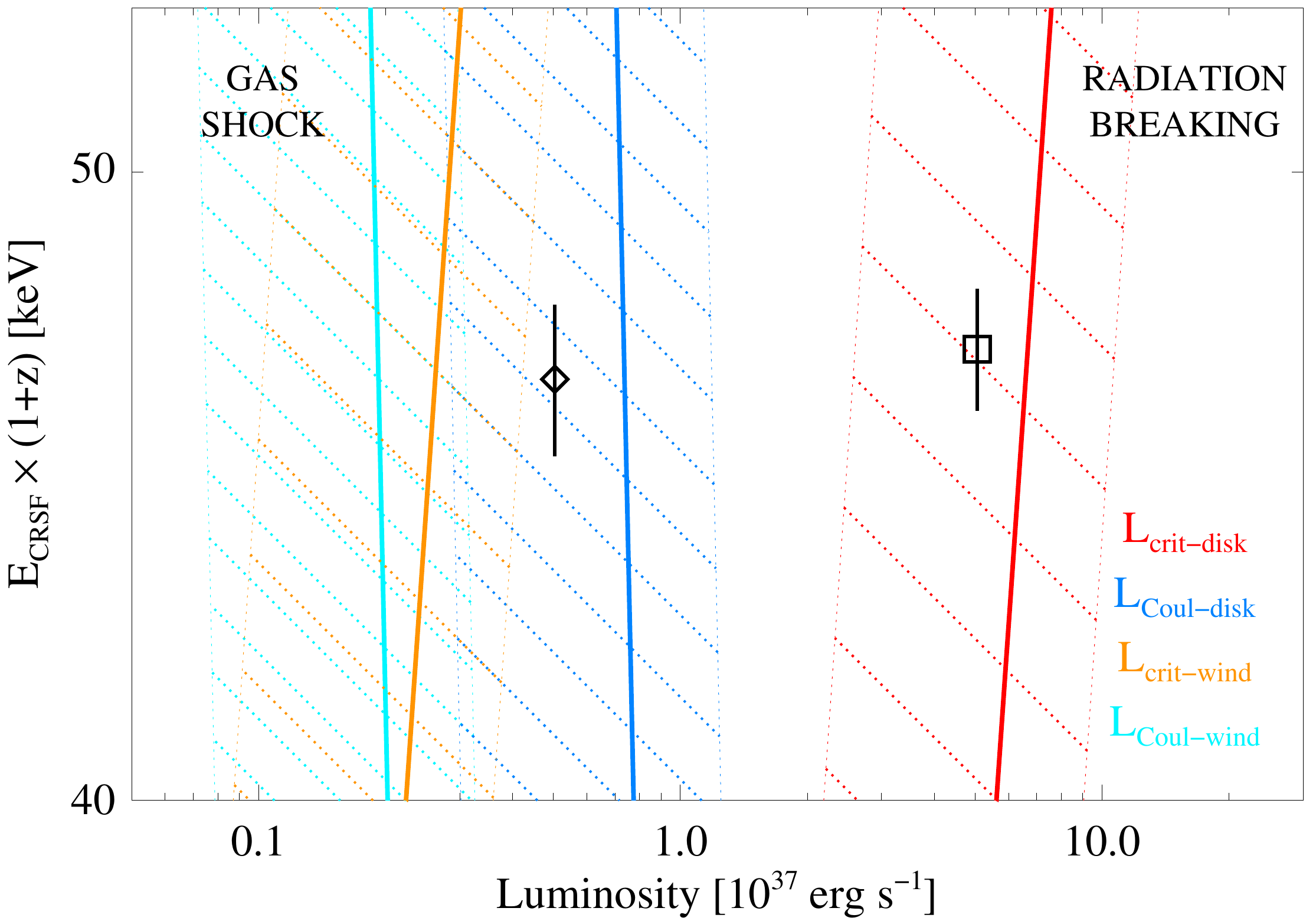}
  \caption{Relationship between the intrinsic CRSF energy and the
    luminosity of XTE\,J1946+274. The results of the spectral fits
    from this work (diamond) and \citet{heindl:01} (square) are shown
    with respect to the Coulomb and critical luminosities
    \citep[equations 32 and 45 in][]{becker:12} of a neutron star with
    a standard mass and radius for the cases of disk (dark blue and
    red solid lines) and wind accretion (light blue and orange solid
    lines). The hashed luminosity ranges account for the uncertainty
    of the distance measurement.}\label{fig:crsf_lum}
\end{figure}

It has recently been recognized that different types of correlations
between the energy of the CRSF $E_\mathrm{CRSF}$ and the X-ray
luminosity $L_\mathrm{X}$ are observed for accreting pulsars, probably
reflecting different accretion states \citep{staubert:07}. Studying
these correlations allows us to derive constraints on the physical
conditions in the accretion column. \citet{becker:12} presented a
model of the different accretion regimes and of how the height (i.e.,
the $B$-field and therefore $E_\mathrm{CRSF}$) of the region in the
accretion column where the CRSF is produced changes with luminosity
for the different regimes \citep[see also][]{mushtukov:15}: For
supercritical sources ($L_\mathrm{X}\gtrsim L_\mathrm{crit}$)
radiation pressure in a radiative shock in the accretion column is the
dominant decelerator for the material inside the accretion column. A
source in this regime is expected to show a negative
$E_\mathrm{CRSF}$-$L_\mathrm{X}$ correlation, as observed for
\objectname{V\,0332+53} \citep{mowlavi:06}. For moderately subcritical
sources ($L_\mathrm{X}\lesssim L_\mathrm{crit}$) the
radiation-dominated shock causes the initial deceleration, followed by
Coulomb interactions below the shock which bring the matter to a stop
on the neutron star surface. Subcritical sources in this regime are
expected to show a positive $E_\mathrm{CRSF}$-$L_\mathrm{X}$
correlation, as observed for \objectname{Her~X-1} or
\objectname{GX~304$-$1} \citep{staubert:07,klochkov:12}. The expected
relationship at even lower luminosities ($L_\mathrm{X}\lesssim
L_\mathrm{Coul}$), where the radiative shock and Coulomb interactions
disappear, and the matter falls through a gas-mediated shock before
hitting the stellar surface, is less clear. \objectname{A\,0535+26},
for example, is a low-luminosity source that does not show any changes
of $E_\mathrm{CRSF}$ in pulse averaged spectra with luminosity
\citep[][but see \citealt{mullerd:13} and
  \citealt{sartore:15}]{caballero:07}.

Where does XTE\,J1946+274 fit into this picture? In
Figure~\ref{fig:crsf_lum} we show the Coulomb luminosity
$L_\mathrm{Coul}$ and the critical luminosity $L_\mathrm{crit}$ for a
range of $B$-fields, i.e., cyclotron line energies
\citep[after][]{becker:12}, separating the different accretion
regimes. These luminosities depend among other things on the accretion
geometry outside of the Afv\'en sphere, two cases are presented: disk
and wind accretion. Overplotted are the gravitational redshift
corrected cyclotron line energies and 3--60\,keV luminosities from
\citet{heindl:01} and from our \textsl{Suzaku} analysis. We calculated
the \textsl{Suzaku} luminosity using the unabsorbed flux measurement
from the FDCUT III spectral fit. The CRSF energy is consistent within
errors between 1998 and 2010, implying that the height of the CRSF
emission region is similar for both observations.  The luminosities,
while both moderate, span a range larger than observed for any other
moderate luminosity pulsar (e.g., Her~X-1 or GX~304$-$1) fall in the
transition region between low and high luminosity pulsars. In the case
of disk accretion, the default assumption for Be-systems, both
luminosities are consistent with subcritical accretion, with the 1998
\textsl{RXTE} measurement at $L_\mathrm{X}\lesssim L_\mathrm{crit}$
and the \textsl{Suzaku} measurement at $L_\mathrm{X}\lesssim
L_\mathrm{Coul}$. Taking the uncertainties of the distance measurement
into account, the similarity of the cyclotron line energy measurements
is not inconsistent with the \citet{becker:12} picture. In the case of
wind accretion XTE\,J1946+274 would have been supercritical during
both measurements and a negative $E_\mathrm{CRSF}$-$L_\mathrm{X}$
correlation would be expected. Calculating the difference in emission
heights for supercritical accretion following equation~(40) of
\citet{becker:12} and assuming a dipole magnetic field, $\Delta
E_\mathrm{CRSF} \lesssim 1.4$\,keV is expected for the two
luminosities. This is comparable to the uncertainties of the two
$E_\mathrm{CRSF}$ measurements, i.e., though unlikely, we cannot rule
out the presence of such a change. We note that wind accretion has so
far only been discussed as a possibility for explaining the
$E_\mathrm{CRSF}$-$L_\mathrm{X}$ relationship of persistent, non-Be,
low luminosity sources like Vela X-1 and 4U\,1538$-$522
\citep{furst:14,hemphill:14}. We did not include the data from
\citet{muller:12} because the presence of a CRSF at 25\,keV at fluxes
between the 1998 and 2010 extremes is only marginally supported (see
\S\ref{subsec:crsf}). However, a higher emission region at
intermediate fluxes in the Coulomb braking regime (disk accretion) is
again consistent with the \citet{becker:12} picture while it is not
consistent in the supercritical regime (wind accretion). Using the
more precise treatment of the critical luminosity by
\citet{mushtukov:15} is qualitatively in agreement with this picture.

We can calculate the CRSF emission region height for a subcritical
source at which the Coulomb interactions start decelerating the plasma
using equation~(51) of \citet{becker:12}:
\begin{multline}
  h_\mathrm{c} = 1.48\times 10^5 \mathrm{cm} \left(\frac{\lambda}{0.1}\right)^{-1} \left(\frac{\tau_*}{20} \right)\left(\frac{M_\mathrm{NS}}{1.4\,M_\odot} \right)^{19/14} \left(\frac{R_\mathrm{NS}}{10\,\mathrm{km}} \right)^{1/14}\\
  \cdot \left(\frac{B_\mathrm{NS}}{10^{12}\,\mathrm{G}} \right)^{-4/7} \left(\frac{L_\mathrm{X}}{10^{37}\,\mathrm{erg\,s}^{-1}} \right)^{-5/7}        
\label{eq:height}
\end{multline}
where the following parameters are as defined in \citet{becker:12}:
$\lambda=0.1$ describes the disk accretion case, $\tau_*\sim20$ is the
Thomson optical depth in the Coulomb regime, $M_\mathrm{NS} =
1.4\,M_\odot$ and $R_\mathrm{NS}=10$\,km are typical values for the
neutron star mass and radius. We obtained $h_\mathrm{c}=211$\,m for
the emission height using $B_\mathrm{NS} = 3.1 (1+z) \times
10^{12}$\,G with $z = 0.3$ and $L_\mathrm{X} =
5\times10^{37}\,\mathrm{erg}\,\mathrm{s}^{-1}$ \citep{heindl:01}.

The similarity of the observed pulse profiles at low and high fluxes
supports a scenario where no strong changes in the emission geometry
happen over and between outbursts. The 2010 \textsl{RXTE} and
\textsl{Suzaku} pulse profiles of XTE\,J1946+274 are double-peaked
with a deep and a shallow minimum that show weak energy dependence of
the depths (Figure~\ref{fig:pulse_profile}). This structure is
strongly similar to what has been observed by \citet{wilson:03} and
\citet{paul:01} during the 1998 outburst with other instruments at
different luminosities. The source even shows a double-peaked profile
during quiescence as observed by \textsl{Chandra} \citep{arabaci:14}.
Interestingly the $\sim$20--40\,keV pulse profile of A\,0535+26 is
very similar to that of XTE\,J1946+274
\citep{caballero:07,sartore:15}. Modeling the profiles of the 2005
August/September outburst of A\,0535+26, \citet{caballero:11}
determined a possible emission pattern by taking into account the
contribution of each of the two magnetic poles. They assumed a dipole
magnetic field with axisymmetric emission regions. The asymmetry of
the pulse profile minima is explained by a small offset of one of the
emission regions from being antipodal. The profiles for A\,0535+26
were obtained when the source had a luminosity of
$L_{3-50\,\mathrm{keV}}\sim 0.8\times10^{37}$\,erg\,s$^{-1}$
\citep{caballero:11}, i.e., not unlike the lower range observed for
XTE\,J1946+274.

In summary, for XTE\,J1946+274 the stability of the pulse profile
shape, the lack of strong changes of the spectral shape
(\S\ref{discussion:continuum}), and the possibly constant CRSF energy
with luminosity all indicate that there have been no major changes in
the accretion column structure and emission geometry over the broad
range of moderate luminosities covered by observations.

\section{SUMMARY AND CONCLUSIONS}\label{sec:summary}

In this paper we analyzed a 50\,ks \textsl{Suzaku} observation of the
accreting pulsar XTE\,J1946+274 taken at the end of the second
outburst in an outburst series in 2010. We performed a detailed
temporal and spectral analysis and compared our results to data
available from other instruments and outbursts. In the following we
summarize the results of our analysis:
\begin{enumerate}

\item We determined a new orbital solution based on \textsl{Fermi}-GBM
  and other data. Its parameters and possible intrinsic pulse period
  evolutions are listed in Table~\ref{tab:periodfit} and shown in
  Figure~\ref{fig:gbm_profile}.

\item We observed no strong changes between the \textsl{Suzaku}
  spectrum and previously analyzed spectra for different luminosities
  and outbursts.

\item The \textsl{Suzaku} observation allowed us to extend the
  correlation between the continuum X-ray flux and the flux of the
  narrow Fe\,K$\alpha$ line to lower fluxes than observed before.
  Comparing the observed correlation with the theoretically expected
  values for fluorescence emission shows a possible slight elevation
  of the line flux. This could indicate either an overabundance of
  iron, a more complex ionization structure, or a more complex spatial
  structure of the emitting medium than assumed by the simplest model.

\item The \textsl{Suzaku} spectrum shows a feature that can be modeled
  with a cyclotron line component at $35.2^{+1.5}_{-1.3}$\,keV at a
  significance of $2.81\sigma$.

\item The unchanging cyclotron line energy and similar pulse profile
  shape with luminosity between 1998 and 2010 suggest that the source
  does not experience strong changes in emission geometry and that
  XTE\,J1946+274 has been consistently accreting in the subcritical
  regime.

\item There are similarities between XTE\,J1946+274 and A\,0535+26
  regarding their pulse profile structure and a possibly unchanging
  cyclotron energy with luminosity. A more detailed study of these
  similarities could prove useful for better understanding accreting
  X-ray pulsars in Be systems.
\end{enumerate}
XTE\,J1946+274 is rarely in outburst, with its two known episodes of
activity having occurred approximately a decade apart. It remains a
source with many unanswered questions. In particular, monitoring of
possible future outbursts with sensitive instruments such as the ones
on \textsl{NuSTAR} or \textsl{Astro-H} could fill the gap in the
cyclotron line energy versus X-ray luminosity correlation and shed new
light on the accretion mechanism of this source.

\acknowledgments

DMM and KP acknowledge support by \textsl{Suzaku} NASA Guest Observer
grant NNX11AD41G and NASA Astrophysical Data Analysis Program grant
12-ADAP12-0118. We acknowledge funding by the Bundesministerium f\"ur
Wirtschaft und Technologie under Deutsches Zentrum f\"ur Luft- und
Raumfahrt grants 50OR1113 and 50OR1207 and Deutsche
Forschungsgemeinschaft grant WI 1860/11-1. We thank John E. Davis for
the development of the \texttt{SLXfig} module, which was used to
create Figure~\ref{fig:gbm_profile}. MTW and KSW acknowledge support
by the Chief of Naval Research and NASA Astrophysical Data Analysis
Program grant 12-ADAP12-0118. VG acknowledges support by NASA through
the Smithsonian Astrophysical Observatory (SAO) contract SV3-73016 to
MIT for Support of the Chandra X-Ray Center (CXC) and Science
Instruments. CXC is operated by SAO for and on behalf of NASA under
contract NAS8-03060.

Facilities: \facility{\textsl{Suzaku}}, \facility{\textsl{Fermi}},
\facility{\textsl{RXTE}}, \facility{\textsl{Swift}},
\facility{\textsl{INTEGRAL}}.


\begin{thebibliography}{}

\bibitem[\protect\astroncite{{Arnaud}}{1996}]{arnaud:96}
{Arnaud}, K.~A.,  1996,
\newblock in ASP Conf. Ser. 101, Astronomical Data Analysis Software and
  Systems V, ed. G.~H. {Jacoby}, J. {Barnes},  (San Francisco, CA:ASP), ~17

\bibitem[\protect\astroncite{{Becker} et~al.}{2012}]{becker:12}
{Becker}, P.~A., {Klochkov}, D., {Sch{\"o}nherr}, G., et~al.\  2012, \aap, 544,
  A123

\bibitem[\protect\astroncite{{Boldt}}{1987}]{boldt:87}
{Boldt}, E.,  1987,
\newblock in IAU Symp. 124, Observational Cosmology, ed. A. {Hewitt}, G.
  {Burbidge}, L.~Z. {Fang},  (Cambridge: Cambridge Univ. Press), 611

\bibitem[\protect\astroncite{{Caballero} et~al.}{2011}]{caballero:11}
{Caballero}, I., {Kraus}, U., {Santangelo}, A., {Sasaki}, M., \& {Kretschmar},
  P.  2011, \aap, 526, A131

\bibitem[\protect\astroncite{{Caballero} et~al.}{2007}]{caballero:07}
{Caballero}, I., {Kretschmar}, P., {Santangelo}, A., et~al.\  2007, \aap, 465,
  L21

\bibitem[\protect\astroncite{{Caballero} et~al.}{2010}]{caballero:10}
{Caballero}, I., {Pottschmidt}, K., {Bozzo}, E., et~al.\  2010, ATel 2692

\bibitem[\protect\astroncite{{Campana} et~al.}{1999}]{campana:99}
{Campana}, S., {Israel}, G., \& {Stella}, L.  1999, \aap, 352, L91

\bibitem[\protect\astroncite{{Coburn} et~al.}{2002}]{coburn:02}
{Coburn}, W., {Heindl}, W.~A., {Rothschild}, R.~E., et~al.\  2002, \apj, 580,
  394

\bibitem[\protect\astroncite{{Enoto} et~al.}{2008}]{enoto:08a}
{Enoto}, T., {Makishima}, K., {Terada}, Y., et~al.\  2008, \pasj, 60, 57

\bibitem[\protect\astroncite{{Finger}}{2010}]{finger:10}
{Finger}, M.~H.,  2010, ATel 2847

\bibitem[\protect\astroncite{{F{\"u}rst} et~al.}{2014}]{furst:14}
{F{\"u}rst}, F., {Pottschmidt}, K., {Wilms}, J., et~al.\  2014, \apj, 780, 133

\bibitem[\protect\astroncite{{Galloway} et~al.}{2004}]{galloway:04}
{Galloway}, D.~K., {Morgan}, E.~H., \& {Levine}, A.~M.  2004, \apj, 613, 1164

\bibitem[\protect\astroncite{{Ghosh} \& {Lamb}}{1979}]{ghosh:79b}
{Ghosh}, P., \& {Lamb}, F.~K.  1979, \apj, 232, 259

\bibitem[\protect\astroncite{{Heindl} et~al.}{2001}]{heindl:01}
{Heindl}, W.~A., {Coburn}, W., {Gruber}, D.~E., et~al.\  2001, \apjl, 563, L35

\bibitem[\protect\astroncite{{Hemphill} et~al.}{2014}]{hemphill:14}
{Hemphill}, P.~B., {Rothschild}, R.~E., {Markowitz}, A., et~al.\  2014, \apj,
  792, 14

\bibitem[\protect\astroncite{{Illarionov} \& {Sunyaev}}{1975}]{illarionov:75}
{Illarionov}, A.~F., \& {Sunyaev}, R.~A.  1975, \aap, 39, 185

\bibitem[\protect\astroncite{{Inoue}}{1985}]{inoue:85}
{Inoue}, H.,  1985, SSRv, 40, 317

\bibitem[\protect\astroncite{{ISAS/JAXA} \& {X-ray Astrophysics Laboratory
  NASA/Goddard Space Flight Center}}{2013}]{abc_guide}
{ISAS/JAXA}\& {X-ray Astrophysics Laboratory NASA/Goddard Space Flight Center}
  2013,
\newblock The Suzaku Data Reduction Guide, Version~5,
\newblock \url{http://heasarc.gsfc.nasa.gov/docs/suzaku/analysis/abc/}

\bibitem[\protect\astroncite{{ISAS/JAXA} \& {X-ray Astrophysics Laboratory
  NASA/Goddard Space Flight Center}}{2015}]{tech_description}
{ISAS/JAXA}\& {X-ray Astrophysics Laboratory NASA/Goddard Space Flight Center}
  2015,
\newblock The Suzaku Technical Description,
\newblock
  \url{http://heasarc.gsfc.nasa.gov/docs/suzaku/prop$\_$tools/suzaku$\_$td/}

\bibitem[\protect\astroncite{{Kalberla} et~al.}{2005}]{kalberla:05}
{Kalberla}, P.~M.~W., {Burton}, W.~B., {Hartmann}, D., et~al.\  2005, \aap,
  440, 775

\bibitem[\protect\astroncite{{Klochkov} et~al.}{2012}]{klochkov:12}
{Klochkov}, D., {Doroshenko}, V., {Santangelo}, A., et~al.\  2012, \aap, 542,
  L28

\bibitem[\protect\astroncite{{Koyama} et~al.}{2007}]{koyama:07}
{Koyama}, K., {Tsunemi}, H., {Dotani}, T., et~al.\  2007, \pasj, 59, 23

\bibitem[\protect\astroncite{{Krimm} et~al.}{2010}]{krimm:10}
{Krimm}, H.~A., {Barthelmy}, S.~D., {Baumgartner}, W., et~al.\  2010, ATel 2663

\bibitem[\protect\astroncite{{K{\"u}hnel} et~al.}{2013}]{kuehnel:13}
{K{\"u}hnel}, M., {M{\"u}ller}, S., {Kreykenbohm}, I., et~al.\  2013, \aap,
  555, A95

\bibitem[\protect\astroncite{{Leahy} \& {Creighton}}{1993}]{leahy:93}
{Leahy}, D.~A., \& {Creighton}, J.  1993, \mnras, 263, 314

\bibitem[\protect\astroncite{{Leahy} et~al.}{1983}]{leahy:83}
{Leahy}, D.~A., {Elsner}, R.~F., \& {Weisskopf}, M.~C.  1983, \apj, 272, 256

\bibitem[\protect\astroncite{Maeda}{2010a}]{suzaku_memo2010_06}
Maeda, Y.,  2010a,
\newblock Photometry using the XIS-1 data taken with the narrow window modes,
\newblock Document JS-ISAS-SUZAKU-MEMO-2010-06,
  \url{ftp://legacy.gsfc.nasa.gov/suzaku/doc/xis/suzakumemo-2010-06.pdf}

\bibitem[\protect\astroncite{Maeda}{2010b}]{suzaku_memo2010_05}
Maeda, Y.,  2010b,
\newblock A possible flux variation of the XIS data taken after Dec. 18th,
  2009,
\newblock Document JS-ISAS-SUZAKU-MEMO-2010-05,
  \url{ftp://legacy.gsfc.nasa.gov/suzaku/doc/xis/suzakumemo-2010-05.pdf}

\bibitem[\protect\astroncite{Maeda}{2010c}]{suzaku_memo2010_04}
Maeda, Y.,  2010c,
\newblock A possible pointing determination error between Dec. 18th, 2009 and
  June 15th, 2010,
\newblock Document JS-ISAS-SUZAKU-MEMO-2010-04,
  \url{ftp://legacy.gsfc.nasa.gov/suzaku/doc/xis/suzakumemo-2010-04.pdf}

\bibitem[\protect\astroncite{Maeda et~al.}{2008}]{suzaku_memo2008_06}
Maeda, Y., Someya, K., Ishida, M., et~al.\  2008,
\newblock Recent update of the XRT response. III.\ Effective Area,
\newblock Document JS-ISAS-SUZAKU-MEMO-2008-06,
  \url{http://www.astro.isas.jaxa.jp/suzaku/doc/suzakumemo/suzakumemo-2008-06.pdf}

\bibitem[\protect\astroncite{{Maitra} \& {Paul}}{2013}]{maitra:13}
{Maitra}, C., \& {Paul}, B.  2013, \apj, 771, 96

\bibitem[\protect\astroncite{{Marcu} et~al.}{2014}]{marcu:14}
{Marcu}, D.~M., {Pottschmidt}, K., {Gottlieb}, A.~M., et~al.\  2014,
\newblock in Proc. of the 10th INTEGRAL Workshop, A Synergistic View of the
  High Energy Sky PoS(INTEGRAL 2014)065

\bibitem[\protect\astroncite{{Mihara}}{1995}]{mihara:95}
{Mihara}, T.,  1995,
\newblock Ph.D. thesis, Dept.~of Physics, Univ.~of Tokyo

\bibitem[\protect\astroncite{{Mowlavi} et~al.}{2006}]{mowlavi:06}
{Mowlavi}, N., {Kreykenbohm}, I., {Shaw}, S.~E., et~al.\  2006, \aap, 451, 187

\bibitem[\protect\astroncite{{M{\"u}ller} et~al.}{2013a}]{mullerd:13}
{M{\"u}ller}, D., {Klochkov}, D., {Caballero}, I., \& {Santangelo}, A.  2013a,
  \aap, 552, A81

\bibitem[\protect\astroncite{{M{\"u}ller} et~al.}{2013b}]{muller:13}
{M{\"u}ller}, S., {Ferrigno}, C., {K{\"u}hnel}, M., et~al.\  2013b, \aap, 551,
  A6

\bibitem[\protect\astroncite{{M{\"u}ller} et~al.}{2012}]{muller:12}
{M{\"u}ller}, S., {K{\"u}hnel}, M., {Caballero}, I., et~al.\  2012, \aap, 546,
  A125

\bibitem[\protect\astroncite{{Mushtukov} et~al.}{2015}]{mushtukov:15}
{Mushtukov}, A.~A., {Suleimanov}, V.~F., {Tsygankov}, S.~S., \& {Poutanen}, J.
  2015, \mnras, 447, 1847

\bibitem[\protect\astroncite{{Nagase} et~al.}{1986}]{nagase:86}
{Nagase}, F., {Hayakawa}, S., {Sato}, N., {Masai}, K., \& {Inoue}, H.  1986,
  \pasj, 38, 547

\bibitem[\protect\astroncite{{Nowak} et~al.}{2011}]{nowak:12}
{Nowak}, M.~A., {Hanke}, M., {Trowbridge}, S.~N., et~al.\  2011, \apj, 728, 13

\bibitem[\protect\astroncite{{\"Ozbey Arabac{\i}} et~al.}{2014}]{arabaci:14}
{\"Ozbey Arabac{\i}}, M., {Camero-Arranz}, A., {Gutierrez-Soto}, J., et~al.\
  2014, \aap,
\newblock in press

\bibitem[\protect\astroncite{{Paul} et~al.}{2001}]{paul:01}
{Paul}, B., {Agrawal}, P.~C., {Mukerjee}, K., et~al.\  2001, \aap, 370, 529

\bibitem[\protect\astroncite{{Reig} \& {Nespoli}}{2013}]{reig:13}
{Reig}, P., \& {Nespoli}, E.  2013, \aap, 551, A1

\bibitem[\protect\astroncite{{Sartore} et~al.}{2015}]{sartore:15}
{Sartore}, N., {Jourdain}, E., \& {Roques}, J.-P.  2015, \apj, 806, 193

\bibitem[\protect\astroncite{{Schwarzenberg-Czerny}}{1989}]{schwarzenberg-czerny:89a}
{Schwarzenberg-Czerny}, A.,  1989, \mnras, 241, 153

\bibitem[\protect\astroncite{{Silaj} et~al.}{2010}]{silaj:10}
{Silaj}, J., {Jones}, C.~E., {Tycner}, C., {Sigut}, T.~A.~A., \& {Smith}, A.~D.
   2010, 187, 228

\bibitem[\protect\astroncite{{Smith} \& {Takeshima}}{1998}]{smith:98}
{Smith}, D.~A., \& {Takeshima}, T.  1998, ATel 36

\bibitem[\protect\astroncite{{Staubert} et~al.}{2007}]{staubert:07}
{Staubert}, R., {Shakura}, N.~I., {Postnov}, K., et~al.\  2007, \aap, 465, L25

\bibitem[\protect\astroncite{{Suchy} et~al.}{2008}]{suchy:08}
{Suchy}, S., {Pottschmidt}, K., {Wilms}, J., et~al.\  2008, \apj, 675, 1487

\bibitem[\protect\astroncite{{Sugizaki} et~al.}{2015}]{sugizaki:15}
{Sugizaki}, M., {Yamamoto}, T., {Mihara}, T., {Nakajima}, M., \& {Makishima},
  K.  2015, \pasj,
\newblock in press (arXiv: 1504.04895)

\bibitem[\protect\astroncite{{Takahashi} et~al.}{2007}]{takahashi:07}
{Takahashi}, T., {Abe}, K., {Endo}, M., et~al.\  2007, \pasj, 59, 35

\bibitem[\protect\astroncite{{Tanaka}}{1986}]{tanaka:86}
{Tanaka}, Y.,  1986,
\newblock in IAU Colloq. 89, Radiation Hydrodynamics in Stars and Compact
  Objects, ed. D. {Mihalas}, K.-H.~A. {Winkler},  (Berlin: Springer),  198

\bibitem[\protect\astroncite{{Titarchuk}}{1994}]{titarchuck:94}
{Titarchuk}, L.,  1994, \apj, 434, 570

\bibitem[\protect\astroncite{{Torrej{\'o}n} et~al.}{2010}]{torrejon:10}
{Torrej{\'o}n}, J.~M., {Schulz}, N.~S., {Nowak}, M.~A., \& {Kallman}, T.~R.
  2010, \apj, 715, 947

\bibitem[\protect\astroncite{Tsujimoto et~al.}{2010}]{suzaku_memo2010_10}
Tsujimoto, M., et~al.\  2010,
\newblock Anomaly of XIS0 in June 2009,
\newblock Document JS-ISAS-SUZAKU-MEMO-2010-01,
  \url{ftp://legacy.gsfc.nasa.gov/suzaku/doc/xis/suzakumemo-2010-01.pdf}

\bibitem[\protect\astroncite{{Verner} \& {Yakovlev}}{1995}]{verner:95}
{Verner}, D.~A., \& {Yakovlev}, D.~G.  1995, \aaps, 109, 125

\bibitem[\protect\astroncite{{Verrecchia} et~al.}{2002}]{verrecchia:02}
{Verrecchia}, F., {Israel}, G.~L., {Negueruela}, I., et~al.\  2002, \aap, 393,
  983

\bibitem[\protect\astroncite{{Wilms} et~al.}{2000}]{wilms:00}
{Wilms}, J., {Allen}, A., \& {McCray}, R.  2000, \apj, 542, 914

\bibitem[\protect\astroncite{{Wilson} et~al.}{2003}]{wilson:03}
{Wilson}, C.~A., {Finger}, M.~H., {Coe}, M.~J., \& {Negueruela}, I.  2003,
  \apj, 584, 996

\bibitem[\protect\astroncite{{Wilson} et~al.}{1998}]{wilson:98}
{Wilson}, C.~A., {Finger}, M.~H., {Wilson}, R.~B., \& {Scott}, D.~M.  1998,
  \iaucirc, 7014

\end{thebibliography}

\end{document}